# Temperature- and charge carrier density-dependent electronic response in methylammonium lead iodide

Jiacheng Wang[#,1,2], Jungmin Park[3], Lei Gao[3], Lucia Di Virgilio[3], Sheng Qu[3], Heejae Kim[3,4], Hai I. Wang[3], Li-Lin Wu[1,6], Wen Zeng[1,2], Mischa Bonn[3*], Zefeng Ren[*,1,2], Jaco J. Geuchies[3,5#*]

1. State Key Laboratory of Molecular Reaction Dynamics, Dalian Institute of Chemical Physics, Chinese Academy of Sciences, 457 Zhongshan Road, Dalian 116023, P. R. China

2. University of Chinese Academy of Sciences, 19A Yuquan Road, Beijing 100049, P.R. China

3. Max Planck Institute for Polymer Research, 55128 Mainz, Germany.

4. Department of Physics, Pohang University of Science and Technology, 37673, Pohang, Korea.

5. Leiden Institute of Chemistry, Leiden University, Einsteinweg 55, 2333CC, Leiden, the Netherlands.

6. School of Physics, Xidian University, Xi'an 710071, P. R. China

* Corresponding author # These authors contributed equally

**ABSTRACT**

Understanding carrier dynamics in photoexcited metal-halide perovskites is key for opto-electronic devices such as solar cells (low carrier densities) and lasers (high carrier densities). Trapping processes at low carrier densities and many-body recombination at high densities can significantly alter the dynamics of photoexcited carriers. Combining optical-pump/THz probe and transient absorption spectroscopy we examine carrier responses over a wide density range ($10^{14}$-$10^{19}$ cm$^{-3}$) and temperatures (78-315K) in the prototypical methylammonium lead iodide perovskite. At densities below ~$10^{15}$ cm$^{-3}$ (room temperature, sunlight conditions), fast carrier trapping at shallow trap states occurs within a few picoseconds. As excited carrier densities increase, trapping saturates, and the carrier response stabilizes, lasting up to hundreds of picoseconds at densities around ~$10^{17}$ cm$^{-3}$. Above $10^{18}$ cm$^{-3}$ a Mott transition sets in: overlapping polaron wavefunctions lead to ultrafast annihilation through an Auger recombination process occurring over a few picoseconds. We map out trap-dominated, direct recombination-dominated, and Mott-dominated density regimes from 78-315 K, ultimately enabling the construction of an electronic "phase diagram". These findings clarify carrier behavior across operational conditions, aiding material optimization for optoelectronics operating in the low (e.g. photovoltaics) and high (e.g. laser) carrier density regimes.

**INTRODUCTION**

Metal-halide perovskite materials are proposed for use in various technologies as a light absorber (e.g., in photodetectors and photovoltaics) or light emitter (e.g., LEDs and lasers), each requiring distinct carrier densities. Even without external injection, methylammonium lead iodide (MAPI) intrinsically hosts charge carriers at densities below $10^{12}$ cm$^{-3}$.[1] As a solar cell material, MAPI operates at carrier densities around $10^{14}$ cm$^{-3}$,[2] where shallow traps dominate carrier dynamics[3]. LEDs, reliant on carrier injection via electronic contacts, operate at intermediate densities around $10^{12}$-$10^{16}$ cm$^{-3}$.[4] For MAPI to be viable for lasing technologies, carrier densities over $10^{18}$ cm$^{-3}$ are required to reach population inversion.[5,6] This wide range of carrier densities, along with the diverse applications envisioned, underscores the need to understand carrier dynamics at different carrier densities across photoexcitation fluences. In addition to hosting extra charge carriers — introduced through electrical injection, chemical doping, or photoexcitation — the perovskite lattice encompasses a multitude of defects[7], each with its own density,[8–17] which can localize or trap charges.

Many reports extensively discuss various types of traps — both deep and shallow, though not always clearly defined — and their associated densities. In polycrystalline lead-halide films, reported defect densities range from $10^{15}$ to $10^{16}$ cm$^{-3}$ [13,14,18–25], while single crystals typically show much lower bulk defect densities, around $10^{12}$ cm$^{-3}$ or lower[13,15,22,26,27]. Recent work by Yuan et al. highlights that in both polycrystalline films and full device architectures of MAPI, shallow traps predominantly influence the time-resolved photoluminescence signals[28], with energy levels ranging from 50 to 130 meV away from the nearest band. Furthermore, the literature suggests the existence of distinctively different carrier density regimes where various recombination processes are predominant, occurring both at very low carrier densities[3,29] and high carrier densities[2,30–35]. This puts into question the concept of 'intrinsic' behavior after impulsive photoexcitation of the perovskite materials.

Here, we systematically study the photoinduced carrier response at densities spanning five orders of magnitude ($10^{14}$-$10^{19}$ cm$^{-3}$) at temperatures between 78-315 K in the prototypical MAPI perovskite. We use highly sensitive transient absorption spectroscopy (TAS) and optical-pump/THz probe (OPTP) spectroscopy to map the carrier dynamics after impulsive photoexcitation. In our TAS experiments, probing the carrier dynamics at $10^{14}$-$10^{15}$ cm$^{-3}$, we observe a fast decay of the photoinduced signal, caused by localization of carriers in shallow traps in several ps. Upon increasing the density to $10^{16}$ cm$^{-3}$, these shallow traps become occupied, and the absorption bleach remains unchanged over hundreds of picoseconds. We extend the density range to $10^{19}$ cm$^{-3}$ using OPTP, and, after an initial flat response of the time-dependent OPTP signal with increasing density, we see the onset of a fast decay over tens of ps and saturation of the carrier density, indicative of a Mott polaron transition around $10^{18}$ cm$^{-3}$. We map out these carrier responses from 78 to 315 K and construct a 'phase diagram', in which we include deep trap densities from literature and a model for predicting optical gain thresholds in MAPI.

**RESULTS AND DISCUSSION**

We synthesized a MAPI thin film on a water-free silica substrate using an established spincoat-and-annealing protocol[36,37]. The resulting polycrystalline MAPI has a preferential orientation with a [110] zone axis, as shown by room-temperature X-ray diffraction measurements (see Figure S1), and a thickness of approximately 300 nm. We sealed the film with epoxy resin in between two water-free glass substrates inside a nitrogen-purged glovebox, to prevent sample degradation under ambient conditions[38]. Metal-halide perovskite materials host a wealth of both shallow and deep traps, which introduce levels inside the bandgap[7,10,12,28]. Furthermore, due to the polycrystalline nature of the thin film, defects at interfaces, such as grain boundaries, cannot be avoided[39].

To follow the evolution of the carrier dynamics as a function of photogenerated carrier density and temperature and construct an electronic 'response diagram', we combine OPTP spectroscopy and highly sensitive TAS. This combination allows us to probe the electronic response over a range of photoexcitation densities spanning five orders of magnitude ($10^{14}$-$10^{19}$ cm$^{-3}$). Briefly, in OPTP experiments, the perovskite is first excited by a 50-fs optical pump pulse, which photoexcites electrons from the valence to the conduction band. After a controlled time delay, a single-cycle THz probe pulse, generated by optical rectification in a ZnTe(110) crystal and with an envelope duration of about 1 ps, interacts with the photoexcited electrons, which attenuates the THz field. The attenuation is a direct measure of the photoconductivity of the sample (given by the sum of products of electron and hole density and their respective mobilities). The transmitted electric field of the THz pulse is detected by an 800 nm sampling pulse via electro-optic sampling in another ZnTe(110) crystal. In TAS experiments, we used a recently developed technique based on a balanced detection scheme, allowing for an exceptional sensitivity ($\Delta T/T$) of ~$10^{-7}$. This high sensitivity enabled us to probe the carrier dynamics in MAPI at carrier densities as low as $10^{14}$ cm$^{-3}$. For the TAS measurements, we used a fs fiber laser (~260 fs pulse duration) to measure the dynamics from fs to ns, and the combination of a ns diode laser and a fs fiber laser in the ns-μs time scale. We used a pump wavelength of about 515 nm in both the TAS and OPTP experiments to enable direct comparison.

For both TAS and OPTP experiments, the density $N$ is inferred from the incident photon fluence and the complex refractive index at the pump wavelength[38] (resulting in a fraction of reflected photons and a decay of the density over the absorption length, see methods section in the SI). Additionally, the photogenerated density is corrected for the photon-to-charge quantum yield, which is determined to be ~30% in the tetragonal phase and ~55% in the orthorhombic phase, which was determined independently from the plasma frequency obtained from fits to the THz conductivity spectra (see Figure S2). As electrons and holes in MAPI have similar effective masses, both contribute nearly equally to the measured photoconductivity.

There are two types of phase transitions in MAPI over the temperature and carrier density range reported here, as shown in Figure 1(a). Around 162 K, MAPI undergoes a structural phase transition from a tetragonal structure (I4/mcm) at higher temperatures, to orthorhombic (Pnma) crystal structure at lower temperatures[40,41], shown in Figure 1(a). In addition, there are also distinct electronic "phases", depending on the carrier density: at high densities, where polaron wavefunctions start to overlap, there is a Mott transition in which polarons rapidly annihilate over tens of picoseconds[32,35]. Here, we measured TAS at carrier densities from $3\times10^{14}$ to $4\times10^{16}$ cm$^{-3}$, and OPTP from $5\times10^{16}$ to $1\times10^{19}$ cm$^{-3}$.

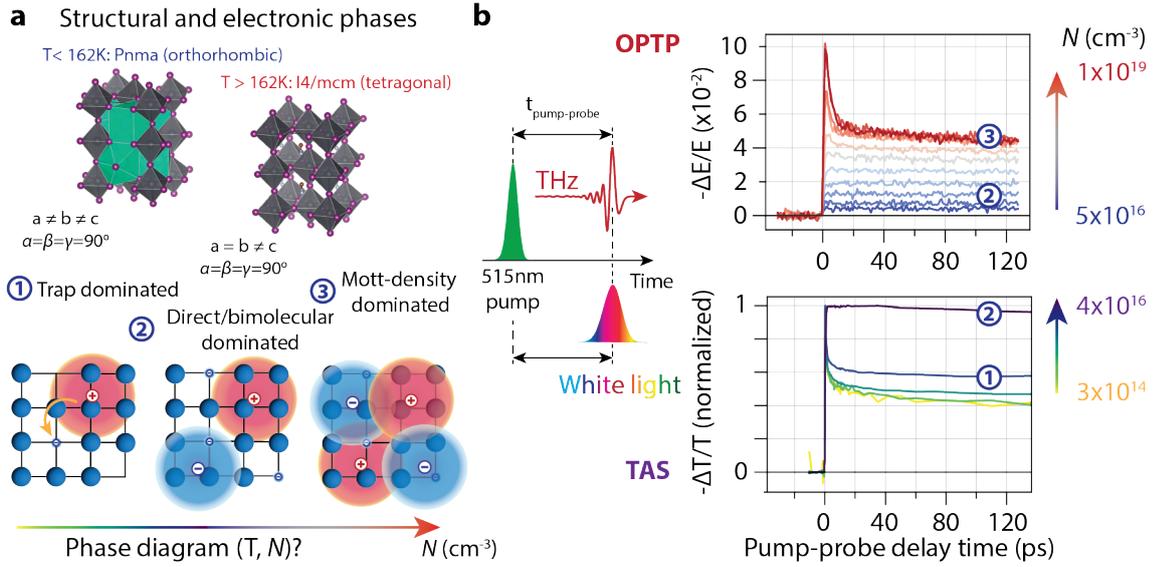

**Figure 1: Structural and electronic phases in methylammonium lead iodide. (a)** MAPI adopts two crystal structures over the temperature range we have investigated; above 162 K it is in the tetragonal (I4/mcm) phase, whereas below it is in the orthorhombic crystal phase. Furthermore, we will show throughout the paper that the dynamics of photoexcited carriers are dominated by (1) trap-assisted recombination (up to $10^{15}$ cm$^{-3}$), (2) direct bimolecular recombination ($10^{15}$-$10^{18}$ cm$^{-3}$), and (3) above the Mott density (> $10^{18}$ cm$^{-3}$), fast polaron annihilation. It is important to note that the trap density is not a material-specific property, but depends on the sample quality, whereas the Mott density is an intrinsic material property. **(b)** The combination of THz and TA pump-probe spectroscopy allows us to probe photoexcited electrons in perovskites over a range of densities spanning five orders of magnitude. THz photons serve as a probe for higher densities ($5\times10^{16} - 1\times10^{19}$ cm$^{-3}$), whereas a white-light probe pulse in transient absorption spectroscopy was used as a probe at low densities ($3\times10^{14} - 4\times10^{16}$ cm$^{-3}$). The electronic phases from panel **(a)** are indicated in the pump-probe transients.

Figure 1(b) shows an example of data recorded at a temperature of 100 K after photoexcitation at 515 nm. The bottom panel shows TAS data in the low-to-medium density range [range 1 and 2 in Figure 1(a)]. At low carrier densities, monomolecular recombination (trap-assisted recombination) dominates carrier recombination. In addition, there is a fast decay over a few ps of the band-edge bleach signal, which we will show indicates the loss of either electrons or holes to shallow trap states. As the carrier density is increased, this fast decay disappears, and the band-edge bleach becomes constant over the first 130 ps, indicating that the fraction of trapped carriers relative to the total photoexcited carrier density decreases and becomes negligible. The decay, mainly through a bimolecular recombination process, and the hot phonon bottleneck effect act together, causing the bleaching signal to remain almost unchanged over the first hundred picoseconds. The top panel shows OPTP data in the medium-to-high carrier density range. Initially, at lower pump fluences, the signal is constant over the first 140 ps. As we increase the pump fluence, the instantaneous photoconductivity increases, but decays fast over the first 10 ps to a constant value, to a density corresponding to the Mott density, which will be discussed later. We start the discussion at low photoexcited carrier densities.

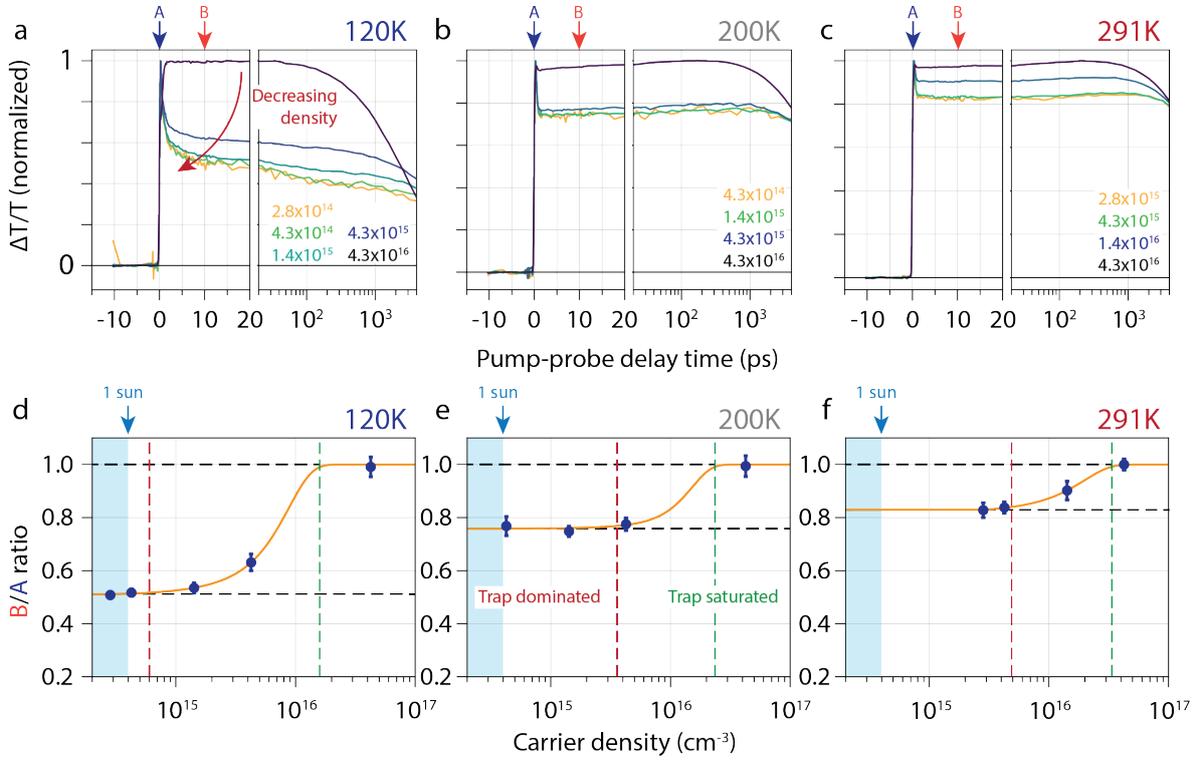

Figure 2: Temperature- and carrier-density-dependent TAS dynamics at different temperatures. (a-c) Density-dependent TA traces at $N < 10^{16}$ cm$^{-3}$ at 120K (a), 200K (b) and 291K (c). At low carrier densities, there is a fast decay of the ΔT/T signal, which we attribute to fast carrier localization into shallow trap states, which get saturated at higher carrier densities. Note that we reach a true linear photoinduced response at the lowest carrier densities at each temperature. The corresponding carrier densities are indicated in the figures. (d-f) Ratio of ΔT/T at 10 ps, $B$, divided by the instantaneous ΔT, $A$, as a function of carrier density at (d) 120K, (e) 200K and (f) 291K. The carrier densities at which we observe a true linear response are indicated by vertical red lines and the carrier

In the TAS measurements, we initially performed spectral measurements at varying temperatures to characterize the ground-state bleach (GSB) feature, see Figure S3. The spectral position of the GSB changes across the tetragonal-to-orthorhombic phase transition in MAPI at 160 K. Figure 2 shows the temperature- and carrier density-dependent TAS measurements at 120 (a), 200 (b), and 291K (c). The data for all temperatures can be found in Figure S4. Reducing the photoexcited carrier density (to $\sim 10^{14}$-$10^{15}$ cm$^{-3}$) causes the normalized TA curves to converge at all temperatures. The nature of trapping is such that a maximum fraction of charge carriers recombine at low carrier densities[3]. We captured this quantitatively by plotting the ratio of ΔT/T at 10 ps, $B$, after the initial fast decay, over the instantaneous ΔT/T, $A$, which are shown in Figures 2(d-f) and S5. Saturation of the B/A ratio at low carrier densities indicates entry into the linear response region, where trap-assisted recombination dominates, lasting tens of nanoseconds, and direct electron-hole recombination and Auger processes are negligible. This region mirrors the intrinsic carrier dynamics of any particular perovskite material with a given trap density under solar illumination ($\sim 10^{14}$ cm$^{-3}$)[3,29]. Across various sample types and preparation methods, a shallow-trapping fraction of tens of percent of the carriers at low temperatures and low carrier densities is typical and seemingly inevitable. Note that the maximum carrier density to enter the linear region depends on temperature, decreasing from $4.3 \times 10^{15}$ cm$^{-3}$ at 291 K to $4.3 \times 10^{14}$ cm$^{-3}$ at 120 K, but the general trend remains consistent.

The reduction of the GSB in the linear response range, which displays a rapid decay within a few picoseconds, is attributed to the trapping of carriers into shallow trap states[3]. As the temperature decreases, a greater proportion of carriers become trapped, because the reduced thermal energy limits their ability to detrap from shallow defect levels. Traps that are shallow at high temperature can act as deep traps at low temperatures, which is also reflected by the decreasing amplitude of the B/A ratio as a function of carrier density for increasing temperatures (see Figure S6). This results from the balance between rapid trapping and reduced thermally activated detrapping. As the pump fluence is increased, the relative amplitude of this initial fast decay becomes vanishingly small, indicating that the fraction of trapped carriers relative to the total photoexcited carrier density decreases, and can even be considered negligible, as these shallow trap states become saturated. At higher carrier densities, $10^{16}$-$10^{17}$ cm$^{-3}$, the carrier dynamics are dominated by direct- or bimolecular recombination. The decay, mainly through a bimolecular recombination process, has a longer lifetime, causing the bleaching signal to remain almost unchanged over hundreds of picoseconds (see Figure S7).

As shown in Figure S8-9 and table S2, the carrier lifetime at 100 K and 120 K extends over several hundred nanoseconds, whereas at higher temperatures, it lasts about tens of nanoseconds. Interestingly, the carrier lifetimes of the orthorhombic structure at low temperatures (100 and 120 K) are nearly identical. Similarly, the carrier lifetimes of the tetragonal structure at high temperatures (170, 200, 220, 250, 270 and 291 K) are also similar, suggesting that the lifetimes are primarily crystal-phase-dependent.

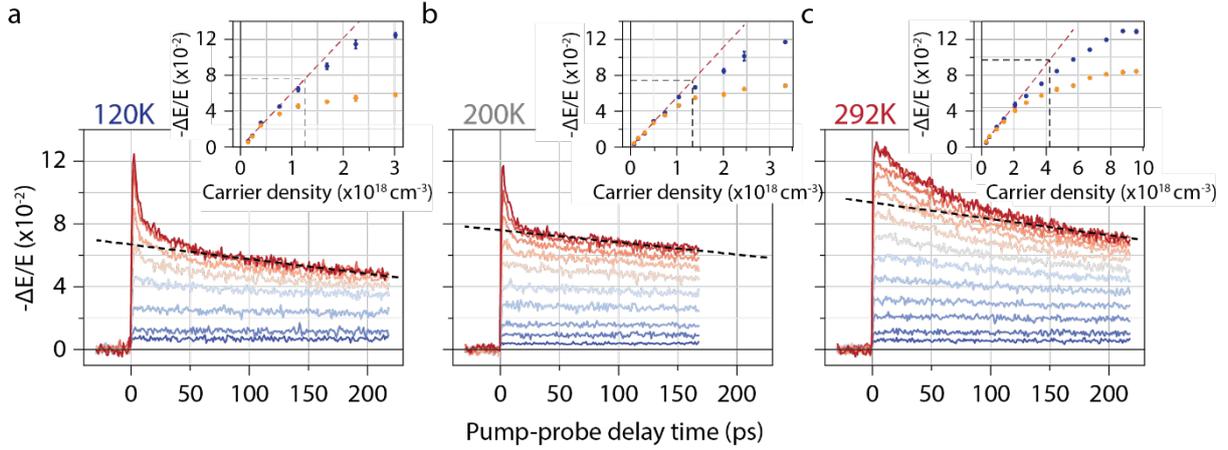

Figure 3: Temperature- and carrier density-dependent polaron decay dynamics measured in the carrier density range of $10^{16}$-$10^{19}$ cm$^{-3}$ with OPTP. (a-c) Density dependent OPTP traces at $N > 10^{16}$ cm$^{-3}$ at (a) 120K, (b) 200K and (c) 292K. As the photogenerated density increases, the polaron wavefunctions start to overlap and they annihilate until the density reaches the Mott density. The insets show the instantaneous -ΔE/E signal (blue datapoints) and late-time -ΔE/E (yellow datapoints), which are used to determine the Mott density (black dashed lines). The red dashed line is a linear fit through the low-fluence data.

We next discuss the carrier dynamics in the medium-to-high carrier density range from $10^{16}$ to $10^{19}$ cm$^{-3}$. Similar to our earlier work[32,35], various features appear in the OPTP transients at low temperatures shown in Figure 3(a): at low carrier densities ($N < 10^{18}$ cm$^{-3}$), the amplitude of the photoconductivity increases linearly with excitation density and shows a modest decay over the pump-probe delay window in our experimental setup (~1 ns, see Figures S10 and S11). The low density ensures that the overlap between polaron wavefunctions is small and there is limited fast bimolecular recombination (i.e., polaron-polaron annihilation). At higher densities, $N > 10^{18}$ cm$^{-3}$, the peak photoconductivity increases sublinearly with $N$, and decays rapidly within the first tens of ps to a constant level. This behavior is summarized by plotting the peak photoconductivity (blue data points) and the photoconductivity at late pump-probe delay times (200 ps, yellow data points) in the inset of Figure 3(a), which reveals the presence of a critical density, $N_{Mott}$. At excitation densities exceeding $N_{Mott}$, the peak photoconductivity increases sub-linearly with $N$, while the photoconductivity at later times reaches a plateau. To accurately determine this critical density, we extrapolate the photoconductivity signal from OPTP traces at high carrier densities at late times back to a pump-probe delay of zero, as indicated by the black dashed line. We utilize the linear relationship established for the photoconductivity at low pump fluences to determine $N_{Mott}$, which is found to be $9.02\pm0.06\times10^{17}$ cm$^{-3}$ at 120 K. Similar behavior is observed at temperatures above 162 K, as illustrated in Figures 3(b) and 3(c). In these cases, where MAPI adopts a tetragonal crystal structure, the photoconductivity exhibits minimal decay at densities below $5\times10^{17}$ cm$^{-3}$. At higher densities, a rapid decay occurs within the first tens of picoseconds, followed by a saturation of the photoconductivity above $N_{Mott}$. The extracted Mott densities are $1.86\pm0.08\times10^{18}$ cm$^{-3}$ and $2.9\pm0.2\times10^{18}$ cm$^{-3}$, at 200 and 292 K, respectively. This is consistent with the formation of large polarons, which screen carriers from defects and other charge carriers[42–47], as we will show below.

Combining the results from both TAS and OPTP over a wide temperature range allowed us to construct "phase diagram" for the electronic response presented in Figure 4. This diagram categorizes the data across the temperature spectrum, distinguishing between the tetragonal and orthorhombic phases of MAPI above and below 160 K, respectively. At densities < $10^{15}$ cm$^{-3}$, fast carrier localization into shallow trap states dominates the photoinduced response. We have extracted the true linear-reponse density, where trap-assisted recombination is dominant, and the shallow-trap-saturated density from the TAS measurements, which are shown as red and green data points, respectively. Direct recombination dominates for intermediate densities spanning $10^{15}$-$10^{17}$ cm$^{-3}$, and photoexcited electrons and holes recombine over a timescale exceeding 1 ns. At densities above $10^{18}$ cm$^{-3}$, the Mott polaron density, fast polaron-polaron annihilation (via an Auger-type mechanism) occurs over a few tens of ps, eventually stabilizing at the Mott density. The Mott densities, obtained by OPTP, are shown as blue datapoints and vary from $1.34\pm0.06\times10^{18}$ cm$^{-3}$ at 78K to $4.5\pm0.1\times10^{18}$ cm$^{-3}$ at 315 K. We can use these obtained Mott densities to estimate the polaron radii as a function of temperature (Figure S12).

The polaron radius decreases from approximately 6.5 nm at 78 K to about 4 nm at 315 K. This trend suggests the formation of large polarons that extend over multiple unit cells, in line with Feynman's polaron theory[32,35,45,46,48]. The depth profile of the density of deep trap levels in MAPI, was determined by Ni et al.[13], and varies from $1.8\times10^{11}$ cm$^{-3}$ in the bulk to $3.1\times10^{12}$ cm$^{-3}$ at the surface and interfaces in MAPI (shown by the purple shaded area).

These results demonstrate that distinct carrier dynamics can be observed at different photogenerated carrier densities, and also emphasize the importance of carefully reporting the carrier densities before concluding the nature of rapidly decaying signals and ascribing them to exact physical effects. This is illustrated by the similar transient signatures of polaron-polaron annihilation through Auger recombination at high densities, and fast trapping of carriers into shallow defect states at low densities, both occurring on the few-to-tens of ps timescales but which have opposite dependencies on carrier density (i.e., incoming photon fluence).

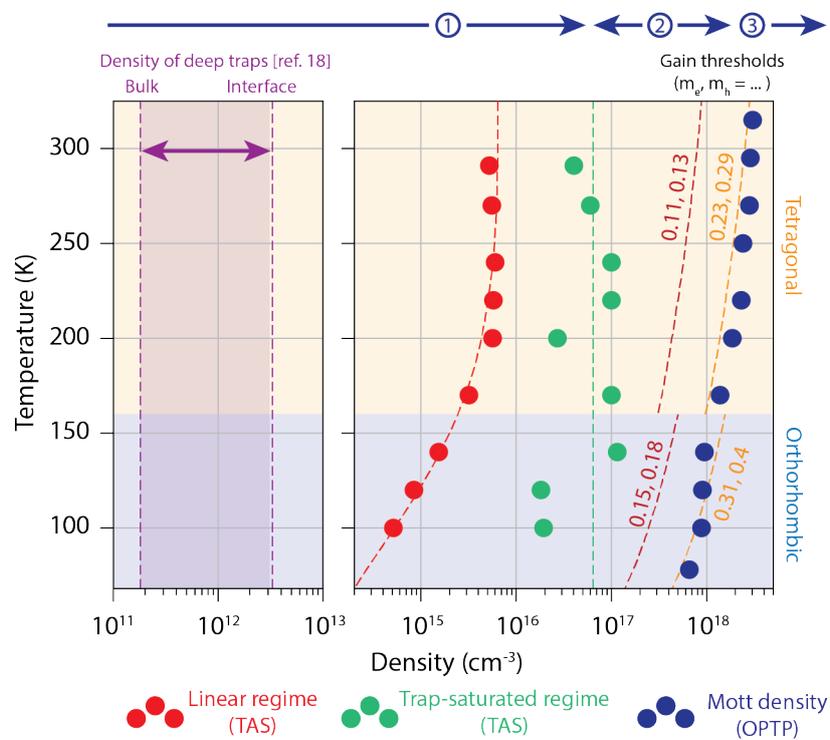
13

**Figure 4: Temperature-dependent electronic "phase diagram" of MAPI spanning eight orders of magnitude in carrier density.** The electronic "phase diagram", was obtained by combining the TAS and OPTP data: the carrier densities corresponding to the linear- and trap-saturated regime were obtained by TAS, the density of deep traps at the interface and the bulk are taken from ref. [12], the Mott density is obtained from OPTP spectroscopy. Gain thresholds for various effective masses in the tetragonal and orthorhombic crystal phases are calculated as explained in the text. The three numbered ranges at the top refer to the electronic density regimes displayed in Figure 1(a).

It is crucial to note that *trap densities are indicative of material quality*, and hence depend on the specific preparation conditions of perovskite thin films. In contrast, the *Mott density is an intrinsic property of the material,* which remains constant regardless of the synthesis details but can be altered by changing the crystal structure or chemical composition.

A possible application for perovskite-based materials at high carrier densities is as a gain medium for lasers. Amplified stimulated emission and lasing have been demonstrated in thin-film MAPI[5,49,50], but concerns have also been raised over data interpretation[51]. In order to reach population inversion, the first requirement for light amplification, the difference in quasi-Fermi-level-splitting for electrons and holes has to be larger than the bandgap[5,52–54]. Using the effective masses for the electrons and holes ($m_e$ = 0.11 $m_0$, $m_h$ = 0.13 $m_0$, with $m_0$ the rest mass of an electron)[55,56] (see SI for the full model description), we estimate that a carrier density of $8\times10^{17}$ cm$^{-3}$ is required for population inversion at room temperature, slightly below the experimentally observed Mott density of $4.2\pm0.2\times10^{18}$ cm$^{-3}$. Over the entire range of experimentally measured temperatures, indicated by orange dashed lines in Figure 4, we observe a notable jump in gain thresholds around 162 K, coinciding with the structural phase transition. In the orthorhombic phase, the carrier effective mass is roughly 36% higher ($m_e$ = 0.15 $m_0$, $m_h$ = 0.18 $m_0$)[57–59]. This increase in effective masses leads to a higher electronic density of states (DOS), necessitating a greater density of photoexcited carriers to achieve population inversion. However, a recent study using angle-resolved photoemission spectroscopy has determined even larger effective masses for electrons and holes: $m_e$, $m_h$ = 0.23 $m_0$, 0.29 $m_0$[60]. The resulting gain thresholds vs. temperature, shown as orange dashed lines, have been scaled in the orthorhombic phase of MAPI using the same factor applied to other reported effective masses. Interestingly, these gain thresholds align closely with the corresponding Mott densities.

Note that across the entire temperature range, using these effective masses, the density required for population inversion aligns with or is below the Mott density. This is encouraging for the development of perovskite materials as a gain medium for lasing applications. Additionally, effects such as bandgap renormalization and strong electron-phonon interactions—both prevalent in MAPI[61,62]—can further lower the estimated gain threshold. These interactions can shift stimulated emission out of the absorbing part of the spectrum, enhancing the material's lasing potential.

**CONCLUSION**

By combining ultrasensitive transient absorption and optical-pump/THz probe spectroscopy, we have captured the ultrafast response of photoexcited carriers in the prototypical perovskite methylammonium lead iodide across carrier densities ranging from $10^{14}$ to $10^{19}$ cm$^{-3}$. A comprehensive analysis allowed us to construct an electronic 'phase diagram' for temperatures ranging from 78 to 315 K. Our findings reveal that at densities below ~$10^{15}$ cm$^{-3}$, there is fast trapping of photoexcited carriers within one-to-several ps into shallow traps, revealed by a rapid decay of the TAS signals. At densities up to $10^{18}$ cm$^{-3}$, the Mott density, the 'shallow-trap-saturated regime' is probed in which the dynamics are constant over the first ns and increase linearly with fluence. Above the Mott density, polaron wavefunctions begin to overlap, causing rapid annihilation of photoexcited polarons within tens of picoseconds, and their size shows the formation of large polarons, consistent with the temperature dependence predicted by Feymann's polaron theory. We demonstrate that the Mott densities exceed both experimentally determined and predicted thresholds for light amplification, underscoring the potential of MAPI as a gain medium. Accurate determination of photoexcited carrier densities is important when reporting ultrafast spectroscopic data, not just for perovskite-based materials, as the transient signals of excited electrons can vary, and may even reverse its dependence on excitation fluence on picosecond timescales, depending on the density range that is probed experimentally.

**Supporting Information available.** Sample preparation and additional steady-state characterization, description of the THz and TAS spectroscopic setups, additional TAS and OPTP data, modelling the threshold for optical gain.

**Competing financial interests.** The authors declare no competing financial interests.

**Corresponding authors.**     Mischa Bonn bonn@mpip-mainz.mpg.de

Zefeng Ren zfren@dipc.ac.cn

Jaco J. Geuchies: j.j.geuchies@lic.leidenuniv.nl

**Acknowledgements.** JJG gratefully acknowledges financial support from the Alexander von Humboldt Foundation. LDV acknowledges the European Union's Horizon 2020 research and innovation program under the Marie Sklodowska-Curie grant No 811284 (UHMob). We thank all members of the THz group of the molecular spectroscopy department at MPIP for many fruitful discussions. This work and JP were supported by the National Research Foundation of Korea (NRF) grant, a Korea-Germany Junior Research Fellowship Program, funded by the Korea government (Ministry of Science and ICT) (No. RS-2023-00254343). ZR, JCW and LLW thank the support by the Strategic Priority Research Program of the Chinese Academy of Sciences (No. XDB0970300).

# REFERENCES

(1) Peña-Camargo, F.; Thiesbrummel, J.; Hempel, H.; Musiienko, A.; Le Corre, V. M.; Diekmann, J.; Warby, J.; Unold, T.; Lang, F.; Neher, D.; Stolterfoht, M. Revealing the Doping Density in Perovskite Solar Cells and Its Impact on Device Performance. *Applied Physics Reviews* **2022**, *9* (2), 021409. https://doi.org/10.1063/5.0085286.

(2) Johnston, M. B.; Herz, L. M. Hybrid Perovskites for Photovoltaics: Charge-Carrier Recombination, Diffusion, and Radiative Efficiencies. *Acc. Chem. Res.* **2016**, *49* (1), 146–154. https://doi.org/10.1021/acs.accounts.5b00411.

(3) Li, B.-H.; Li, H.; Di, H.; Xuan, Z.; Zeng, W.; Wang, J.-C.; Cheng, D.-B.; Zhou, C.; Wang, X.; Zhao, Y.; Zhang, J.; Ren, Z.; Yang, X. Probing the Genuine Carrier Dynamics of Semiconducting Perovskites under Sunlight. *JACS Au* **2023**, *3* (2), 441–448. https://doi.org/10.1021/jacsau.2c00581.

(4) Wang, S.; Cao, Y.; Peng, Q.; Huang, W.; Wang, J. Carrier Dynamics Determines the Optimization Strategies of Perovskite LEDs and PVs. *Research* **2023**, *6*, 0112. https://doi.org/10.34133/research.0112.

(5) Suárez, I.; Juárez-Pérez, E. J.; Chirvony, V. S.; Mora-Seró, I.; Martínez-Pastor, J. P. Mechanisms of Spontaneous and Amplified Spontaneous Emission in ${\mathrm{CH}}_{3}{\mathrm{NH}}_{3}{\mathrm{Pb}\mathrm{I}}_{3}$ Perovskite Thin Films Integrated in an Optical Waveguide. *Phys. Rev. Appl.* **2020**, *13* (6), 064071. https://doi.org/10.1103/PhysRevApplied.13.064071.

(6) Qin, J.; Liu, X.-K.; Yin, C.; Gao, F. Carrier Dynamics and Evaluation of Lasing Actions in Halide Perovskites. *Trends in Chemistry* **2021**, *3* (1), 34–46. https://doi.org/10.1016/j.trechm.2020.10.010.

(7) Mosquera-Lois, I.; Huang, Y.-T.; Lohan, H.; Ye, J.; Walsh, A.; Hoye, R. L. Z. Multifaceted Nature of Defect Tolerance in Halide Perovskites and Emerging Semiconductors. *Nat Rev Chem* **2025**, *9* (5), 287–304. https://doi.org/10.1038/s41570-025-00702-w.

(8) Wolf, N. R.; Jaffe, A.; Slavney, A. H.; Mao, W. L.; Leppert, L.; Karunadasa, H. I. Tuning Defects in a Halide Double Perovskite with Pressure. *J. Am. Chem. Soc.* **2022**, *144* (45), 20763–20772. https://doi.org/10.1021/jacs.2c08607.

(9) Caselli, V. M.; Thieme, J.; Jöbsis, H. J.; Phadke, S. A.; Zhao, J.; Hutter, E. M.; Savenije, T. J. Traps in the Spotlight: How Traps Affect the Charge Carrier Dynamics in Cs2AgBiBr6 Perovskite. *Cell Reports Physical Science* **2022**, 101055. https://doi.org/10.1016/j.xcrp.2022.101055.

(10) Li, Y.; Jia, Z.; Yang, Y.; Yao, F.; Liu, Y.; Lin, Q. Shallow Traps-Induced Ultra-Long Lifetime of Metal Halide Perovskites Probed with Light-Biased Time-Resolved Microwave Conductivity. *Applied Physics Reviews* **2023**, *10* (1), 011406. https://doi.org/10.1063/5.0129883.

(11) Righetto, M.; Lim, S. S.; Giovanni, D.; Lim, J. W. M.; Zhang, Q.; Ramesh, S.; Tay, Y. K. E.; Sum, T. C. Hot Carriers Perspective on the Nature of Traps in Perovskites. *Nature Communications* **2020**, *11* (1), 2712. https://doi.org/10.1038/s41467-020-16463-7.

(12) Jin, H.; Debroye, E.; Keshavarz, M.; G. Scheblykin, I.; J. Roeffaers, M. B.; Hofkens, J.; A. Steele, J. It's a Trap! On the Nature of Localised States and Charge Trapping in Lead Halide Perovskites. *Materials Horizons* **2020**, *7* (2), 397–410. https://doi.org/10.1039/C9MH00500E.

(13) Ni, Z.; Bao, C.; Liu, Y.; Jiang, Q.; Wu, W.-Q.; Chen, S.; Dai, X.; Chen, B.; Hartweg, B.; Yu, Z.; Holman, Z.; Huang, J. Resolving Spatial and Energetic Distributions of Trap States in Metal Halide Perovskite Solar Cells. *Science* **2020**, *367* (6484), 1352–1358. https://doi.org/10.1126/science.aba0893.

(14) Hutter, E. M.; Eperon, G. E.; Stranks, S. D.; Savenije, T. J. Charge Carriers in Planar and Meso-Structured Organic-Inorganic Perovskites: Mobilities, Lifetimes, and Concentrations of Trap States. *Journal of Physical Chemistry Letters* **2015**, *6* (15), 3082–3090. https://doi.org/10.1021/acs.jpclett.5b01361.

(15) Shi, D.; Adinolfi, V.; Comin, R.; Yuan, M.; Alarousu, E.; Buin, A.; Chen, Y.; Hoogland, S.; Rothenberger, A.; Katsiev, K.; Losovyj, Y.; Zhang, X.; Dowben, P. A.; Mohammed, O. F.; Sargent, E. H.; Bakr, O. M. Low Trap-State Density and Long Carrier Diffusion in Organolead Trihalide Perovskite Single Crystals. *Science* **2015**, *347* (6221), 519–522. https://doi.org/10.1126/science.aaa2725.

(16) Tulus; Muscarella, L. A.; Galagan, Y.; Boehme, S. C.; von Hauff, E. Trap Passivation and Suppressed Electrochemical Dynamics in Perovskite Solar Cells with C60 Interlayers. *Electrochimica Acta* **2022**, *433*, 141215. https://doi.org/10.1016/j.electacta.2022.141215.

(17) Siekmann, J.; Ravishankar, S.; Kirchartz, T. Apparent Defect Densities in Halide Perovskite Thin Films and Single Crystals. *ACS Energy Lett.* **2021**, *6* (9), 3244–3251. https://doi.org/10.1021/acsenergylett.1c01449.

(18) Futscher, M. H.; Gangishetty, M. K.; Congreve, D. N.; Ehrler, B. Quantifying Mobile Ions and Electronic Defects in Perovskite-Based Devices with Temperature-Dependent Capacitance Measurements: Frequency vs Time Domain. *The Journal of Chemical Physics* **2020**, *152* (4), 044202. https://doi.org/10.1063/1.5132754.

(19) Son, D.-Y.; Kim, S.-G.; Seo, J.-Y.; Lee, S.-H.; Shin, H.; Lee, D.; Park, N.-G. Universal Approach toward Hysteresis-Free Perovskite Solar Cell via Defect Engineering. *J. Am. Chem. Soc.* **2018**, *140* (4), 1358–1364. https://doi.org/10.1021/jacs.7b10430.

(20) Jiang, L.-L.; Wang, Z.-K.; Li, M.; Zhang, C.-C.; Ye, Q.-Q.; Hu, K.-H.; Lu, D.-Z.; Fang, P.-F.; Liao, L.-S. Passivated Perovskite Crystallization via G-C3N4 for High-Performance Solar Cells. *Advanced Functional Materials* **2018**, *28* (7), 1705875. https://doi.org/10.1002/adfm.201705875.

(21) Wu, Y.; He, Y.; Li, S.; Li, X.; Liu, Y.; Sun, Q.; Cui, Y.; Hao, Y.; Wu, Y. Efficient Inverted Perovskite Solar Cells with Preferential Orientation and Suppressed Defects of Methylammonium Lead Iodide by Introduction of Phenothiazine as Additive. *Journal of Alloys and Compounds* **2020**, *823*, 153717. https://doi.org/10.1016/j.jallcom.2020.153717.

(22) Dong, Q.; Fang, Y.; Shao, Y.; Mulligan, P.; Qiu, J.; Cao, L.; Huang, J. Electron-Hole Diffusion Lengths > 175 Mm in Solution-Grown CH3NH3PbI3 Single Crystals. *Science* **2015**, *347* (6225), 967–970. https://doi.org/10.1126/science.aaa5760.

(23) Zohar, A.; Kulbak, M.; Levine, I.; Hodes, G.; Kahn, A.; Cahen, D. What Limits the Open-Circuit Voltage of Bromide Perovskite-Based Solar Cells? *ACS Energy Lett.* **2019**, *4* (1), 1–7. https://doi.org/10.1021/acsenergylett.8b01920.

(24) Duan, H.-S.; Zhou, H.; Chen, Q.; Sun, P.; Luo, S.; Song, T.-B.; Bob, B.; Yang, Y. The Identification and Characterization of Defect States in Hybrid Organic–Inorganic Perovskite Photovoltaics. *Phys. Chem. Chem. Phys.* **2014**, *17* (1), 112–116. https://doi.org/10.1039/C4CP04479G.

(25) Pan, J.; Chen, Z.; Zhang, T.; Hu, B.; Ning, H.; Meng, Z.; Su, Z.; Nodari, D.; Xu, W.; Min, G.; Chen, M.; Liu, X.; Gasparini, N.; Haque, S. A.; Barnes, P. R. F.; Gao, F.; Bakulin, A. A. Operando Dynamics of Trapped Carriers in Perovskite Solar Cells Observed via Infrared Optical Activation Spectroscopy. *Nat Commun* **2023**, *14* (1), 8000. https://doi.org/10.1038/s41467-023-43852-5.

(26) Adinolfi, V.; Yuan, M.; Comin, R.; Thibau, E. S.; Shi, D.; Saidaminov, M. I.; Kanjanaboos, P.; Kopilovic, D.; Hoogland, S.; Lu, Z.-H.; Bakr, O. M.; Sargent, E. H. The In-Gap Electronic State Spectrum of Methylammonium Lead Iodide Single-Crystal Perovskites. *Advanced Materials* **2016**, *28* (17), 3406–3410. https://doi.org/10.1002/adma.201505162.

(27) Liu, Y.; Zhang, Y.; Zhao, K.; Yang, Z.; Feng, J.; Zhang, X.; Wang, K.; Meng, L.; Ye, H.; Liu, M.; Liu, S. (Frank). A 1300 Mm2 Ultrahigh-Performance Digital Imaging Assembly Using High-Quality Perovskite Single Crystals. *Advanced Materials* **2018**, *30* (29), 1707314. https://doi.org/10.1002/adma.201707314.

(28) Yuan, Y.; Yan, G.; Dreessen, C.; Rudolph, T.; Hülsbeck, M.; Klingebiel, B.; Ye, J.; Rau, U.; Kirchartz, T. Shallow Defects and Variable Photoluminescence Decay Times up to 280 Ms in Triple-Cation Perovskites. *Nat. Mater.* **2024**, 1–7. https://doi.org/10.1038/s41563-023-01771-2.

(29) Li, B.-H.; Di, H.; Li, H.; Wang, J.-C.; Zeng, W.; Cheng, D.-B.; Zhou, C.; Wang, X.; Shi, Y.; Song, J.; Zhao, Y.; Yang, X.; Ren, Z. Unveiling the Intrinsic Photophysics in Quasi-Two-Dimensional

Perovskites. *J. Am. Chem. Soc.* **2024**, *146* (10), 6974–6982. https://doi.org/10.1021/jacs.3c14737.

(30) Herz, L. M. Charge-Carrier Mobilities in Metal Halide Perovskites: Fundamental Mechanisms and Limits. *ACS Energy Letters* **2017**, *2* (7), 1539–1548. https://doi.org/10.1021/acsenergylett.7b00276.

(31) Davies, C. L.; Filip, M. R.; Patel, J. B.; Crothers, T. W.; Verdi, C.; Wright, A. D.; Milot, R. L.; Giustino, F.; Johnston, M. B.; Herz, L. M. Bimolecular Recombination in Methylammonium Lead Triiodide Perovskite Is an Inverse Absorption Process. *Nature Communications 2018 9:1* **2018**, *9* (1), 1–9. https://doi.org/10.1038/s41467-017-02670-2.

(32) Zhang, H.; Debroye, E.; Vina-Bausa, B.; Valli, D.; Fu, S.; Zheng, W.; Di Virgilio, L.; Gao, L.; Frost, J. M.; Walsh, A.; Hofkens, J.; Wang, H. I.; Bonn, M. Stable Mott Polaron State Limits the Charge Density in Lead Halide Perovskites. *ACS Energy Lett.* **2022**, 420–428. https://doi.org/10.1021/acsenergylett.2c01949.

(33) Lim, J.; Kober-Czerny, M.; Lin, Y.-H.; Ball, J. M.; Sakai, N.; Duijnstee, E. A.; Hong, M. J.; Labram, J. G.; Wenger, B.; Snaith, H. J. Long-Range Charge Carrier Mobility in Metal Halide Perovskite Thin-Films and Single Crystals via Transient Photo-Conductivity. *Nat Commun* **2022**, *13* (1), 4201. https://doi.org/10.1038/s41467-022-31569-w.

(34) Guo, Z.; Wan, Y.; Yang, M.; Snaider, J.; Zhu, K.; Huang, L. Long-Range Hot-Carrier Transport in Hybrid Perovskites Visualized by Ultrafast Microscopy. *Science* **2017**, *356* (6333), 59–62. https://doi.org/10.1126/science.aam7744.

(35) Gao, L.; Zhang, H.; Zhang, Y.; Fu, S.; Geuchies, J. J.; Valli, D.; Saha, R. A.; Pradhan, B.; Roeffaers, M.; Debroye, E.; Hofkens, J.; Lu, J.; Ni, Z.; Wang, H. I.; Bonn, M. Tailoring Polaron Dimensions in Lead-Tin Hybrid Perovskites. *Advanced Materials* **2024**, *36* (40), 2406109. https://doi.org/10.1002/adma.202406109.

(36) Yalcinkaya, Y.; Hermes, I. M.; Seewald, T.; Amann-Winkel, K.; Veith, L.; Schmidt-Mende, L.; Weber, S. A. L. Chemical Strain Engineering of MAPbI3 Perovskite Films. **2022**. https://doi.org/10.48550/arxiv.2205.00381.

(37) Zhang, W.; Saliba, M.; Moore, D. T.; Pathak, S. K.; Hörantner, M. T.; Stergiopoulos, T.; Stranks, S. D.; Eperon, G. E.; Alexander-Webber, J. A.; Abate, A.; Sadhanala, A.; Yao, S.; Chen, Y.; Friend, R. H.; Estroff, L. A.; Wiesner, U.; Snaith, H. J. Ultrasmooth Organic–Inorganic Perovskite Thin-Film Formation and Crystallization for Efficient Planar Heterojunction Solar Cells. *Nat Commun* **2015**, *6* (1), 6142. https://doi.org/10.1038/ncomms7142.

(38) Leguy, A. M. A.; Hu, Y.; Campoy-Quiles, M.; Alonso, M. I.; Weber, O. J.; Azarhoosh, P.; van Schilfgaarde, M.; Weller, M. T.; Bein, T.; Nelson, J.; Docampo, P.; Barnes, P. R. F. Reversible Hydration of CH3NH3PbI3 in Films, Single Crystals, and Solar Cells. *Chem. Mater.* **2015**, *27* (9), 3397–3407. https://doi.org/10.1021/acs.chemmater.5b00660.

(39) Kumar, A.; Gupta, S. K.; Dhamaniya, B. P.; Pathak, S. K.; Karak, S. Understanding the Origin of Defect States, Their Nature, and Effects on Metal Halide Perovskite Solar Cells. *Materials Today Energy* **2023**, *37*, 101400. https://doi.org/10.1016/j.mtener.2023.101400.

(40) Sharma, R.; Menahem, M.; Dai, Z.; Gao, L.; Brenner, T. M.; Yadgarov, L.; Zhang, J.; Rakita, Y.; Korobko, R.; Pinkas, I.; Rappe, A. M.; Yaffe, O. Lattice Mode Symmetry Analysis of the Orthorhombic Phase of Methylammonium Lead Iodide Using Polarized Raman. *Phys. Rev. Mater.* **2020**, *4* (5), 051601. https://doi.org/10.1103/PhysRevMaterials.4.051601.

(41) La-O-Vorakiat, C.; Xia, H.; Kadro, J.; Salim, T.; Zhao, D.; Ahmed, T.; Lam, Y. M.; Zhu, J. X.; Marcus, R. A.; Michel-Beyerle, M. E.; Chia, E. E. M. Phonon Mode Transformation Across the Orthohombic-Tetragonal Phase Transition in a Lead Iodide Perovskite CH3NH3PbI3: A Terahertz Time-Domain Spectroscopy Approach. *Journal of Physical Chemistry Letters* **2016**, *7* (1), 1–6. https://doi.org/10.1021/acs.jpclett.5b02223.

(42) Zhang, H.; Debroye, E.; Steele, J. A.; Roeffaers, M. B. J.; Hofkens, J.; Wang, H. I.; Bonn, M. Highly Mobile Large Polarons in Black Phase CsPbI3. *ACS Energy Lett.* **2021**, *6* (2), 568–573. https://doi.org/10.1021/acsenergylett.0c02482.

(43) Zheng, F.; Wang, L. W. Large Polaron Formation and Its Effect on Electron Transport in Hybrid Perovskites. *Energy and Environmental Science* **2019**, *12* (4), 1219–1230. https://doi.org/10.1039/c8ee03369b.

(44) Zhu, X.-Y.; Podzorov, V. Charge Carriers in Hybrid Organic–Inorganic Lead Halide Perovskites Might Be Protected as Large Polarons. *J. Phys. Chem. Lett.* **2015**, *6* (23), 4758–4761. https://doi.org/10.1021/acs.jpclett.5b02462.

(45) Bao, D.; Chang, Q.; Chen, B.; Chen, X.; Sun, H.; Lam, Y. M.; Zhao, D.; Zhu, J.-X.; Chia, E. E. M. Evidence of Polaron Formation in Halide Perovskites via Carrier Effective Mass Measurements. *PRX Energy* **2023**, *2* (1), 013001. https://doi.org/10.1103/PRXEnergy.2.013001.

(46) Bretschneider, S. A.; Ivanov, I.; Wang, H. I.; Miyata, K.; Zhu, X.; Bonn, M.; Bretschneider, S. A.; Ivanov, I.; Wang, H. I.; Bonn, M.; Miyata, K.; Zhu, X. Quantifying Polaron Formation and Charge Carrier Cooling in Lead-Iodide Perovskites. *Advanced Materials* **2018**, *30* (29), 1707312. https://doi.org/10.1002/ADMA.201707312.

(47) Cinquanta, E.; Meggiolaro, D.; Motti, S. G.; Gandini, M.; Alcocer, M. J. P.; Akkerman, Q. A.; Vozzi, C.; Manna, L.; De Angelis, F.; Petrozza, A.; Stagira, S. Ultrafast THz Probe of Photoinduced Polarons in Lead-Halide Perovskites. *Physical Review Letters* **2019**, *122* (16), 166601. https://doi.org/10.1103/PhysRevLett.122.166601.

(48) Martin, B. A. A.; Frost, J. M. Multiple Phonon Modes in Feynman Path-Integral Variational Polaron Mobility. *Phys. Rev. B* **2023**, *107* (11), 115203. https://doi.org/10.1103/PhysRevB.107.115203.

(49) Palmieri, T.; Baldini, E.; Steinhoff, A.; Akrap, A.; Kollár, M.; Horváth, E.; Forró, L.; Jahnke, F.; Chergui, M. Mahan Excitons in Room-Temperature Methylammonium Lead Bromide Perovskites. *Nat Commun* **2020**, *11* (1), 850. https://doi.org/10.1038/s41467-020-14683-5.

(50) Li, Z.; Moon, J.; Gharajeh, A.; Haroldson, R.; Hawkins, R.; Hu, W.; Zakhidov, A.; Gu, Q. Room-Temperature Continuous-Wave Operation of Organometal Halide Perovskite Lasers. *ACS Nano* **2018**, *12* (11), 10968–10976. https://doi.org/10.1021/acsnano.8b04854.

(51) Brenner, P.; Paetzold, U. W.; Turnbull, G. A.; Giebink, N. C.; Samuel, I. D. W.; Lemmer, U.; Howard, I. A. Comment on "Room-Temperature Continuous-Wave Operation of Organometal Halide Perovskite Lasers." *ACS Nano* **2019**, *13* (11), 12257–12258. https://doi.org/10.1021/acsnano.9b00133.

(52) Bernard, M. G. A.; Duraffourg, G. Laser Conditions in Semiconductors. *physica status solidi (b)* **1961**, *1* (7), 699–703. https://doi.org/10.1002/pssb.19610010703.

(53) Yablonovitch, E.; Kane, E. O. Band Structure Engineering of Semiconductor Lasers for Optical Communications. *Journal of Lightwave Technology* **1988**, *6* (8), 1292–1299. https://doi.org/10.1109/50.4133.

(54) Rosencher, E.; Vinter, B. *Optoelectronics*; Piva, P. G., Translator; Cambridge University Press: Cambridge, 2002. https://doi.org/10.1017/CBO9780511754647.

(55) Frohna, K.; Deshpande, T.; Harter, J.; Peng, W.; Barker, B. A.; Neaton, J. B.; Louie, S. G.; Bakr, O. M.; Hsieh, D.; Bernardi, M. Inversion Symmetry and Bulk Rashba Effect in Methylammonium Lead Iodide Perovskite Single Crystals. *Nat Commun* **2018**, *9* (1), 1829. https://doi.org/10.1038/s41467-018-04212-w.

(56) Miyata, A.; Mitioglu, A.; Plochocka, P.; Portugall, O.; Wang, J. T. W.; Stranks, S. D.; Snaith, H. J.; Nicholas, R. J. Direct Measurement of the Exciton Binding Energy and Effective Masses for Charge Carriers in Organic-Inorganic Tri-Halide Perovskites. *Nature Physics* **2015**, *11* (7), 582–587. https://doi.org/10.1038/nphys3357.

(57) Zhong, M.; Zeng, W.; Tang, H.; Wang, L.-X.; Liu, F.-S.; Tang, B.; Liu, Q.-J. Band Structures, Effective Masses and Exciton Binding Energies of Perovskite Polymorphs of CH3NH3PbI3. *Solar Energy* **2019**, *190*, 617–621. https://doi.org/10.1016/j.solener.2019.08.055.

(58) Frost, J. M.; Walsh, A. What Is Moving in Hybrid Halide Perovskite Solar Cells? *Acc. Chem. Res.* **2016**, *49* (3), 528–535. https://doi.org/10.1021/acs.accounts.5b00431.

(59) Zhai, Y.; Wang, K.; Zhang, F.; Xiao, C.; Rose, A. H.; Zhu, K.; Beard, M. C. Individual Electron and Hole Mobilities in Lead-Halide Perovskites Revealed by Noncontact Methods. *ACS Energy Lett.* **2020**, *5* (1), 47–55. https://doi.org/10.1021/acsenergylett.9b02310.

(60) Park, J.; Huh, S.; Choi, Y. W.; Kang, D.; Kim, M.; Kim, D.; Park, S.; Choi, H. J.; Kim, C.; Yi, Y. Visualizing the Low-Energy Electronic Structure of Prototypical Hybrid Halide Perovskite through Clear Band Measurements. *ACS Nano* **2024**, *18* (10), 7570–7579. https://doi.org/10.1021/acsnano.3c12587.

(61) Saidi, W. A.; Poncé, S.; Monserrat, B. Temperature Dependence of the Energy Levels of Methylammonium Lead Iodide Perovskite from First-Principles. *J. Phys. Chem. Lett.* **2016**, *7* (24), 5247–5252. https://doi.org/10.1021/acs.jpclett.6b02560.

(62) Price, M. B.; Butkus, J.; Jellicoe, T. C.; Sadhanala, A.; Briane, A.; Halpert, J. E.; Broch, K.; Hodgkiss, J. M.; Friend, R. H.; Deschler, F. Hot-Carrier Cooling and Photoinduced Refractive Index Changes in Organic–Inorganic Lead Halide Perovskites. *Nat Commun* **2015**, *6* (1), 8420. https://doi.org/10.1038/ncomms9420.

**For table of contents only**

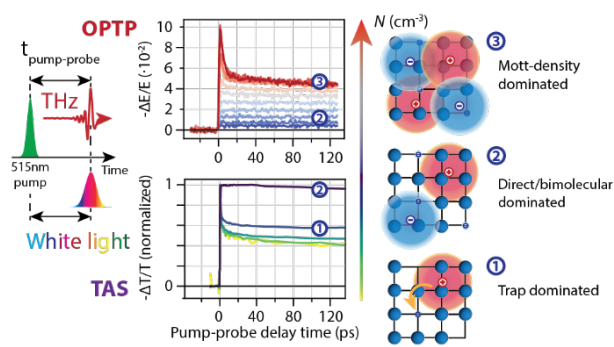

# Supporting information

# Temperature- and charge carrier density-dependent electronic response in methylammonium lead iodide

Jiacheng Wang[#,1,2], Jungmin Park[3], Lei Gao[3], Lucia Di Virgilio[3], Sheng Qu[3], Heejae Kim[3,4], Hai I. Wang[3], Li-Lin Wu[1,6], Wen Zeng[1,2], Mischa Bonn[3*], Zefeng Ren[*,1,2], Jaco J. Geuchies[3,5#*]

1. State Key Laboratory of Molecular Reaction Dynamics, Dalian Institute of Chemical Physics, Chinese Academy of Sciences, 457 Zhongshan Road, Dalian 116023, P. R. China

2. University of Chinese Academy of Sciences, 19A Yuquan Road, Beijing 100049, P.R. China

3. Max Planck Institute for Polymer Research, 55128 Mainz, Germany.

4. Department of Physics, Pohang University of Science and Technology, 37673, Pohang, Korea.

5. Leiden Institute of Chemistry, Leiden University, Einsteinweg 55, 2333CC, Leiden, the Netherlands.

6. School of Physics, Xidian University, Xi'an 710071, P. R. China

# Table of Contents

## Sample preparation

Methylammonium iodide (MAI) was purchased from Greatcell Energy. Lead acetate [(Pb(Ac)$_2$) 98% purity] was purchased from Tokyo Chemical Industry. Anhydrous dimethylformamide (>99.8% purity) was purchased from Sigma Aldrich. Substrates (0.3mm-thick glass) were cleaned by sonication in water with soap for 30 minutes, afterwards rinsed with deionized water three times, and again sonicated in ethanol and acetone (both for 30 minutes). Afterwards, the substrates were dried under a nitrogen flow and subjected to UV-Ozone treatment (FHR UVOH 150 LAB, 250 W) for 20 minutes with an oxygen feeding rate of 1L/min right before spincoating. Film preparation was carried out in a nitrogen purged glovebox. The precursor solution consists out of 477mg (3mmol) MAI and 325.3mg of Pb(Ac)$_2$ (1mmol) in 1 mL DMF. The film preparation was done by depositing 50 µL of the precursor solution onto a cleaned fused silica substrate and spincoating it at 4000 rpm (ramp ± 1000 rpm/s) for one minute. Afterwards the film was left to dry at room temperature for 5 minutes and subsequently annealed on a 100°C hotplate. We have sealed the film with epoxy-resin in-between two water-free glass substrates inside a nitrogen purged glovebox, to prevent sample degradation under ambient conditions.

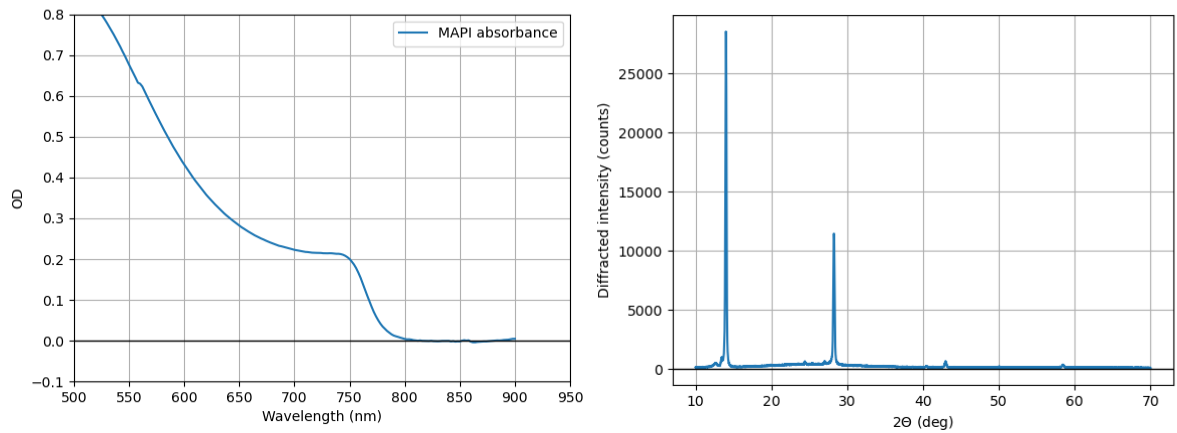

**Figure S1: Absorption spectrum (left) and powder X-ray diffraction pattern of the [110] oriented MAPI film used throughout this work.** The absorption spectrum was measured by placing the sample inside an integrating sphere. X-ray diffraction was performed on a Rigaku Smartlab diffractometer, using copper Kα radiation (1.54 angstrom) in Bragg-Brentano geometry.

## Ultrafast spectroscopic methods

### Optical-pump/THz probe spectroscopy

Here, we describe the technique THz time-domain spectroscopy (TDS) and Optical Pump THz Probe Spectroscopy (OPTP). This technique can probe the photoconductivity of carriers and further retrieve the charge carrier mobility $\mu$.

We use an amplified Ti:sapphire laser (Spitfire Ace, Spectra-Physics) producing pulses with 800 nm central wavelength and ~50 fs pulse duration at 1 kHz repetition rate. The THz field is generated by optical rectification in a ZnTe(110) crystal (thickness 1 mm). The THz detection is based on the electro-optic (also called Pockels) effect in a second ZnTe crystal with 1 mm thickness. We vary the time delay between the THz field and the 800 nm sampling beam with a motorized delay stage (M-605.2DD purchased from Physik Instrument (PI)). The time delay between the optical pump and THz probe pulses is controlled by a second motorized delay stage (M521.DD, Physik Instrument (PI)). The pump-pulse is generated in an optical parametric amplifier (OPA) to convert our 800nm fundamental beam into 515 nm, to match the experimental conditions between the OPTP and TAS measurements, which we filter further using a 515±10 nm bandpass filter (FBH515-10, Thorlabs) directly after the OPA.

## Highly-sensitive transient-absorption spectroscopy

The transient absorption measurements were based on a fiber laser (1035 nm, pulse duration of ~230 fs, 1 MHz repetition rate). The sensitivity level ($\Delta T/T$) of $10^{-7}$ was achieved by a novel technique of combining macropulse and micropulse and using a balanced detector scheme. The output was split into two beams. The first beam (pump) frequency doubled by second harmonic generation. The pump light was modulated at 500 Hz of macropulses by a chopper. The other beam of 1035 nm was focused into a sapphire crystal to generate a broad supercontinuum probe pulse. Afterwards, it was split into two beams by using an achromatic waveplate and a pellicle beam splitter (PBS). One was guided over a monochromator and was collected by a photodiode (PD) as the reference light, and the other was focused and then passed through the sample as the probe light, collimated and dispersed with a monochromator, and finally probed by a second PD, with an integrating 3 nm bandwidth (FWHM) of the probe light. Both PDs were connected to a balanced transimpedance amplifier and then a lock-in amplifier. For the TA measurements at longer pump-probe delay times, a diode laser (516 nm, FWHM 2.2 nm, NPL52C, Thorlabs) was used as the pump, which was synchronized with the fiber laser. Their delay was electronically tuned with a delay generator.

## Determination of the photoexcitation density

The photon density impinging on the sample can be calculated as

$$n_{photon} = \frac{E\lambda}{hc} = \frac{Power\ (W)}{repetition\ rate\ (s^{-1}) \cdot \frac{h \cdot c}{\lambda}\ (J) \cdot \pi r^2\ (cm^2)}$$

A fraction of the incoming light is reflected, $F_R$, at the perovskite-air interface, which can be calculated using the following equations

$$F_A \cdot F_R = (1 - 10^A) \cdot F_R = (1 - 10^A) \cdot \left|\frac{n_1 - n_2}{n_1 + n_2}\right|^2$$

where $n_1$ and $n_2$ are the refractive indexes of the materials at the pump wavelength (515 nm), $A$ is the absorption value at the corresponding wavelength. Here, $n_1 = 1 + 0i$ (for air) and $n_2 = 2.9992 + 0.3523i$ (for MAPbI$_3$)[1]. Since the thickness of the film (~300 nm) is much thicker than the absorption length at the pump wavelength ( $L_{abs,515nm} = \frac{1}{\alpha} = \frac{\lambda}{4\pi \kappa} = \frac{515\ nm}{4\pi\ 0.3523} \approx 116\ nm$ ), we set $F_A = 1$. To calculate the photoexcited carrier density, we have to take into account that the pump beam attenuates over the absorption length in the sample following the Beer-Lambert law, characterized by a decay length of a few tens nanometers from the surface, as

$$I(x) = I_0 e^{-\alpha x}$$

The corresponding absorbed light flux over the absorption length/penetration depth is

$$\left(\frac{1 - e^{-\alpha L}}{\alpha L}\right) \cdot n_{photon}$$

where $\alpha$ is the absorption coefficient, $L \ (= 1/\alpha)$ is the penetration depth at the pump wavelength (or the sample thickness if this is smaller than the penetration depth).

Considering all the parameters, the modified absorbed carrier density can be expressed as:

$$N_{abs} = \frac{Power}{repetition\ rate \cdot \frac{h \cdot c}{\lambda} \cdot \pi r^2 \cdot L} \cdot (1 - 10^A) \cdot \left|\frac{n_1 - n_2}{n_1 + n_2}\right|^2 \cdot \left(\frac{1 - e^{-\alpha L}}{\alpha L}\right)$$

The photogenerated carrier density $N$ in THz spectroscopy (not for TAS) is inferred from the absorbed carrier density, and the photon-to-carrier quantum yield $\Phi$: $N = \Phi N_{abs}$. The absorbance of MAPI at 515 nm is nearly temperature-independent[2], and hence we have used the room temperature refractive index at this wavelength for the calculations of the photoexcited carrier density.

## Photoconductivity spectra and photon-to-carrier quantum yield

We have determined the photo-to-carrier quantum yield $\Phi$ independently, by recording a photoconductivity spectrum at low fluence (and hence carrier density) for the MAPI film we have measured. This was fitted with the Drude-Smith model, in order to extract the plasma frequency, proportional to the carrier density:

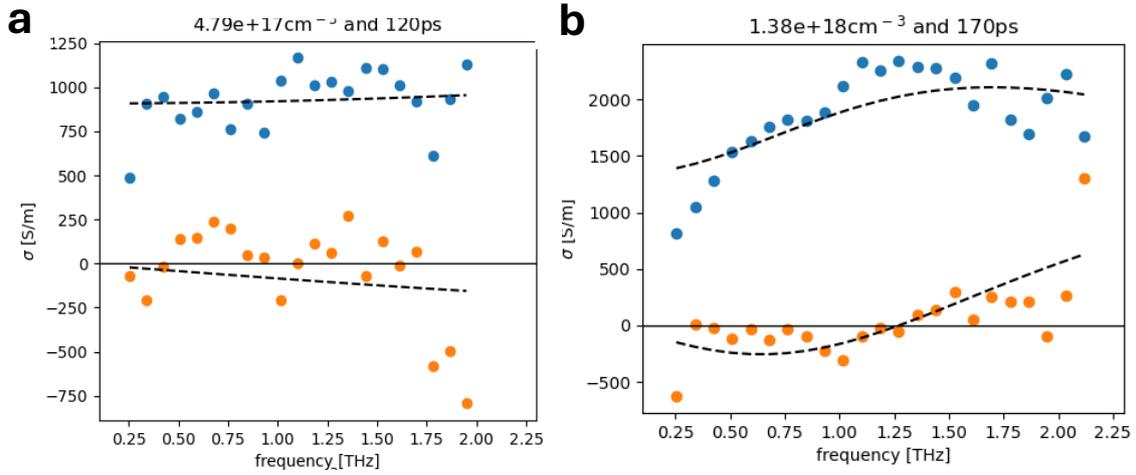

**Figure S2: Photoconductivity spectra at low photoexcited carrier densities and late pump-probe delay times at (a) 78 K and (b) 292 K.** Blue and orange datapoints correspond to the real and the imaginary part of the complex photoconductivity respectively. The dashed lines are a fit to the Drude-Smith model.

$$\Delta\tilde{\sigma}(\omega) = \frac{\varepsilon_0 \omega_p^2 \tau_0}{(1 - i\omega\tau_0)} \cdot \left(1 + \frac{C}{1 - i\omega\tau_0}\right)$$

$\omega_p$ is the plasma frequency, $\varepsilon_0$ is the vacuum permittivity $\tau_0$ is the elastic scattering time, and $C$ is the backscattering constant, a measure for the probability that the charge maintains its initial velocity during scattering ranging from 0 (Drude scattering only, represents full momentum randomized scattering) to −1 (backscattering only). From the fit, the charge carrier density ($N = \omega_p^2 m^* \frac{\varepsilon_0}{e^2}$) can be extracted and the photon-to-carrier quantum yield $\Phi$ can be calculated by comparison with the absorbed photon density. The results of the Drude-Smith fits to the photoconductivity spectra are shown in the table below.

| Temperature | $\tau_0$ (fs) | $\omega_p$ (THz) | $C$ |
|---|---|---|---|
| 78 K | 53±10 | 63±7 | -0.55±0.05 |
| 292 K | 72±4 | 81±3 | -0.67±0.02 |

**Supplementary table 1: Results from the fits of the Drude-Smith model to the photoconductivity spectra shown in Figure S2.**

We obtained photon-to-carrier quantum yields of 0.55 at 78 K (for the orthorhombic phase), and 0.3 at 292 K (for the tetragonal phase).

# Full TAS dataset and analysis

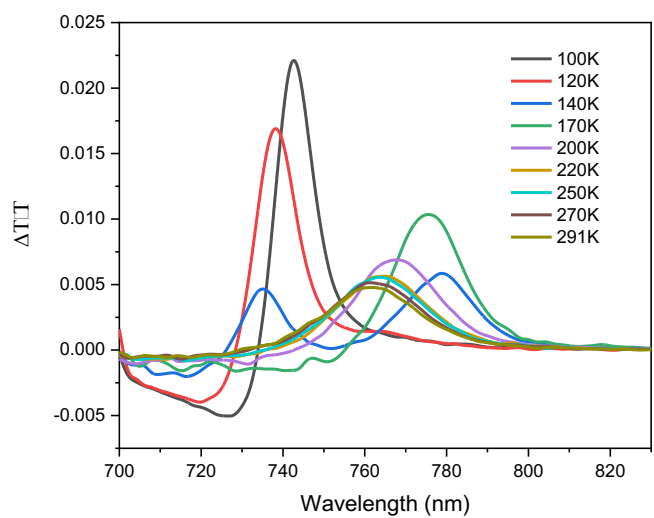

**Figure S3: TA spectra at different temperatures.** Ensemble of TA spectra of MAPbI$_3$ thin films at different temperatures and changes of transmission $\Delta T/T$ were plotted at a pump-probe delay of 1 ns. The pump energy was 10 nJ per pulse with a diameter of 3.5 mm and a wavelength of 517.5 nm. Note that two GSB peaks imply the presence of two phases at 140 K.

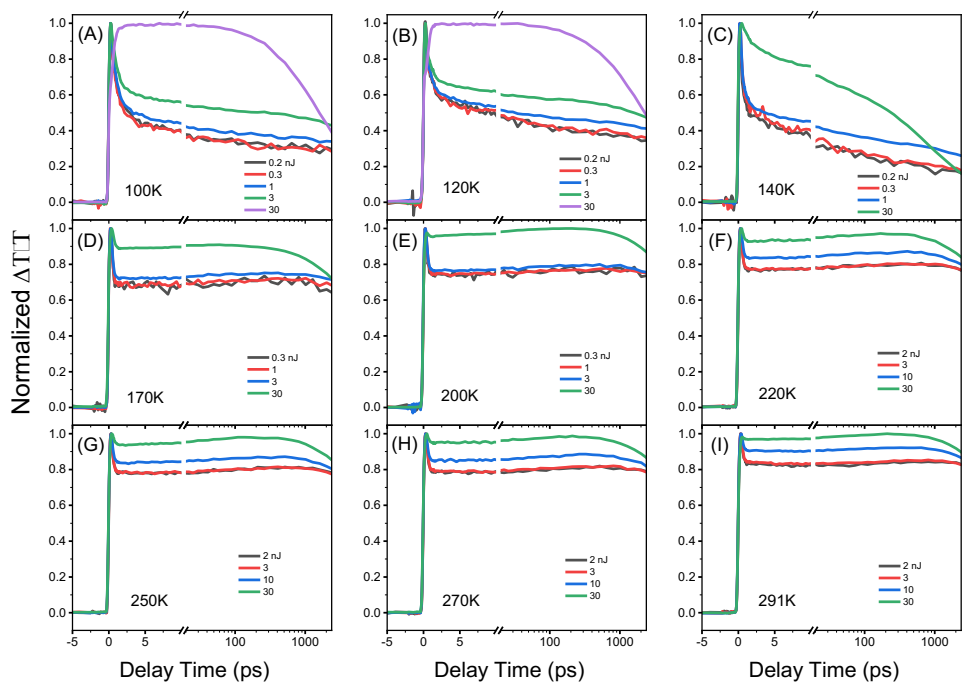

**Figure S4. Carrier density-dependent TA dynamics excited by femtosecond laser pulses at different temperatures.** A 3 mm diameter femtosecond laser with a wavelength of 517.5 nm was used as the pump light. The wavelengths of the bleaching peaks at each temperature obtained from Figure S3 were used as the wavelengths of the probe light (diameter ~0.8 mm) at the corresponding temperatures, respectively.

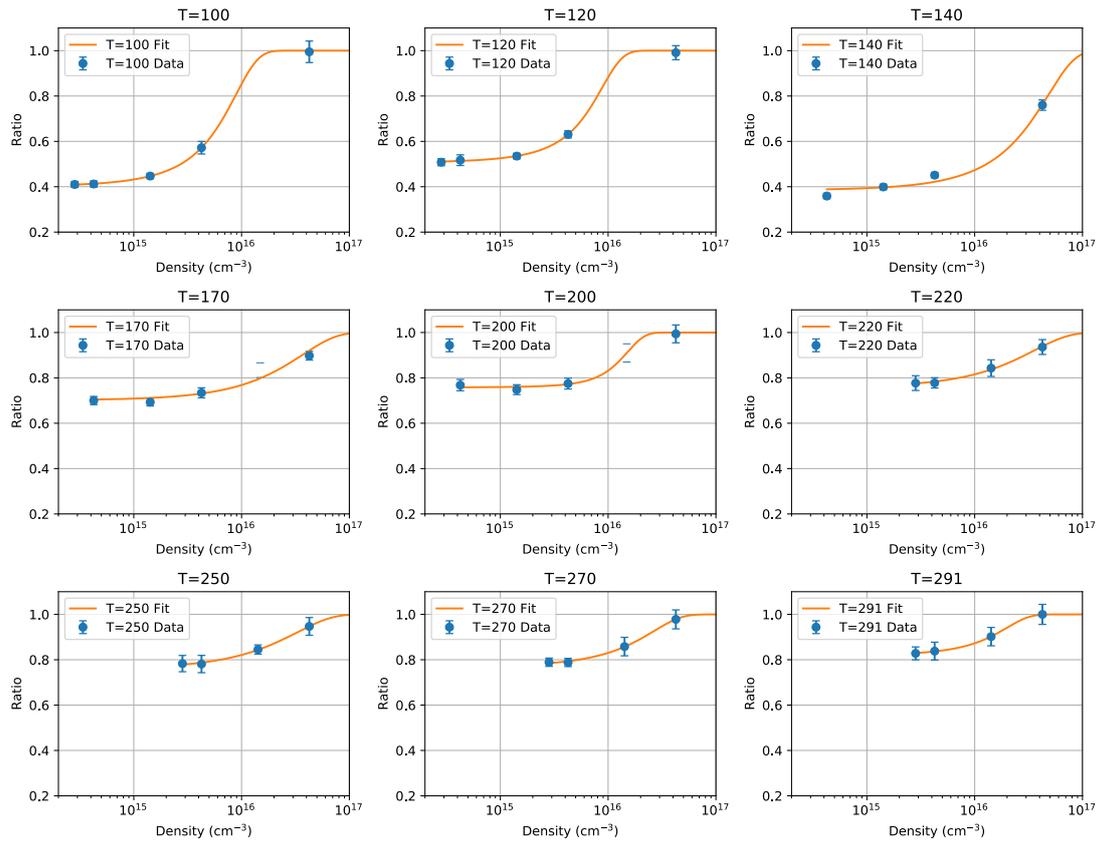

**Figure S5.** *B/A* **ratio (blue datapoints) versus carrier density at all measured temperatures.** The yellow lines are fits to an error function.

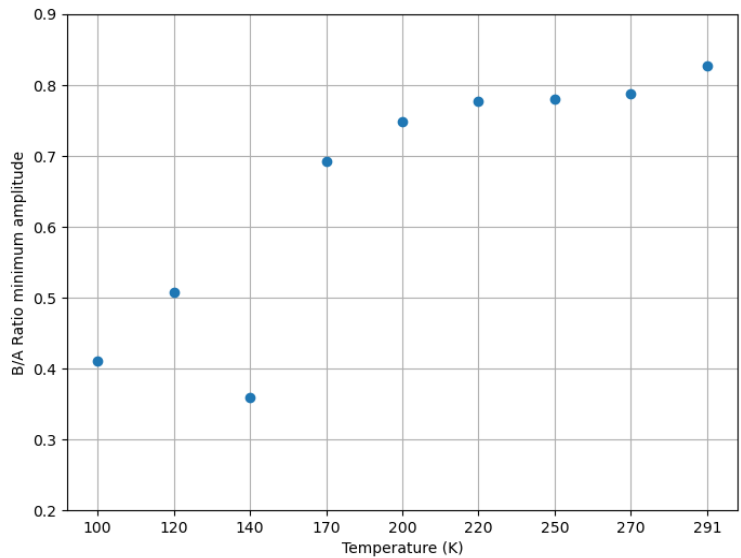

**Figure S6. Amplitude of the *B/A* ratio error function fit versus temperature.** Lower numbers in this case mean larger amplitudes. The deviation at 140K is likely due to the presence of both orthorhombic and tetragonal crystal phases.

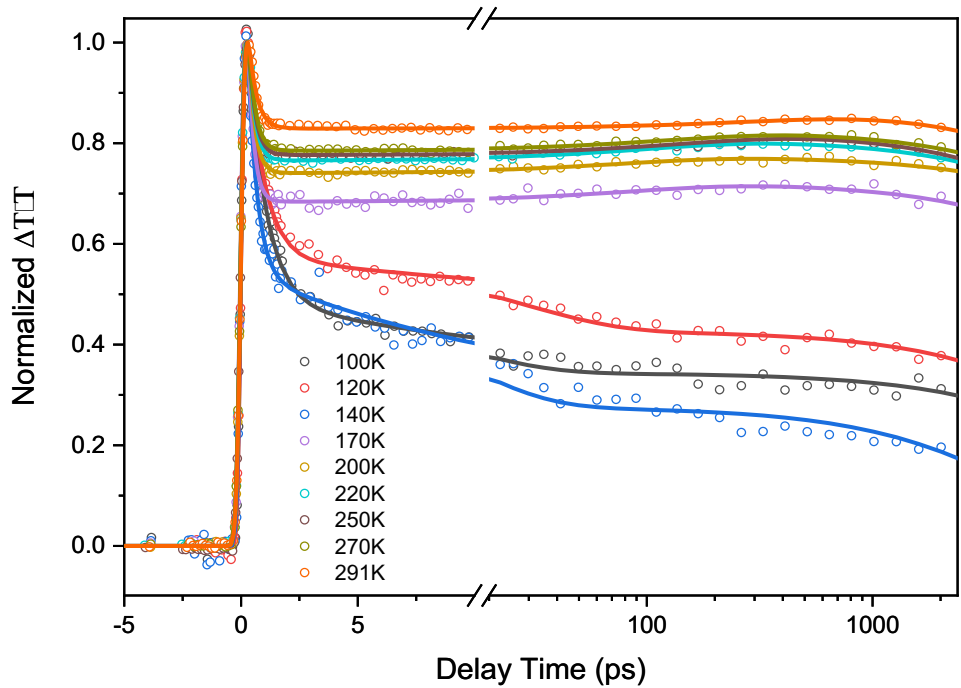

**Figure S7: Comparison of TA dynamics excited by femtosecond laser at different temperatures in the linear response range.** Dots are original data, and lines are fitting results.

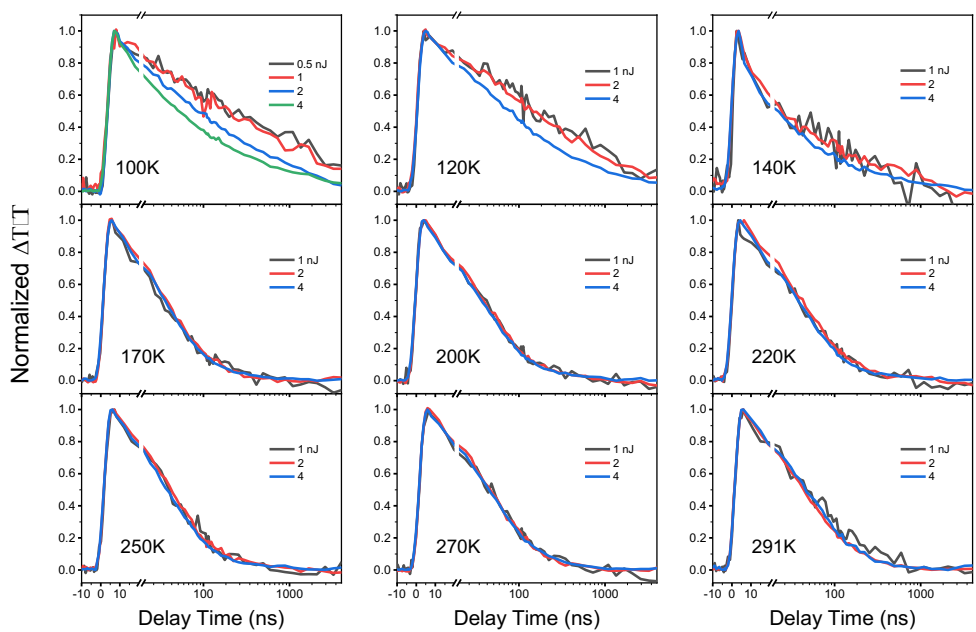

**Figure S8: Carrier density-dependent TA dynamics excited by nanosecond laser at different temperatures.** The pump light is a nanosecond laser with a spot size of about $8 \times 2.5$ mm². The wavelengths of the bleaching peaks at each temperature obtained from Figure S3 were used as the wavelengths of the probe light (light spot diameter about 0.8 mm) at the corresponding temperatures, respectively.

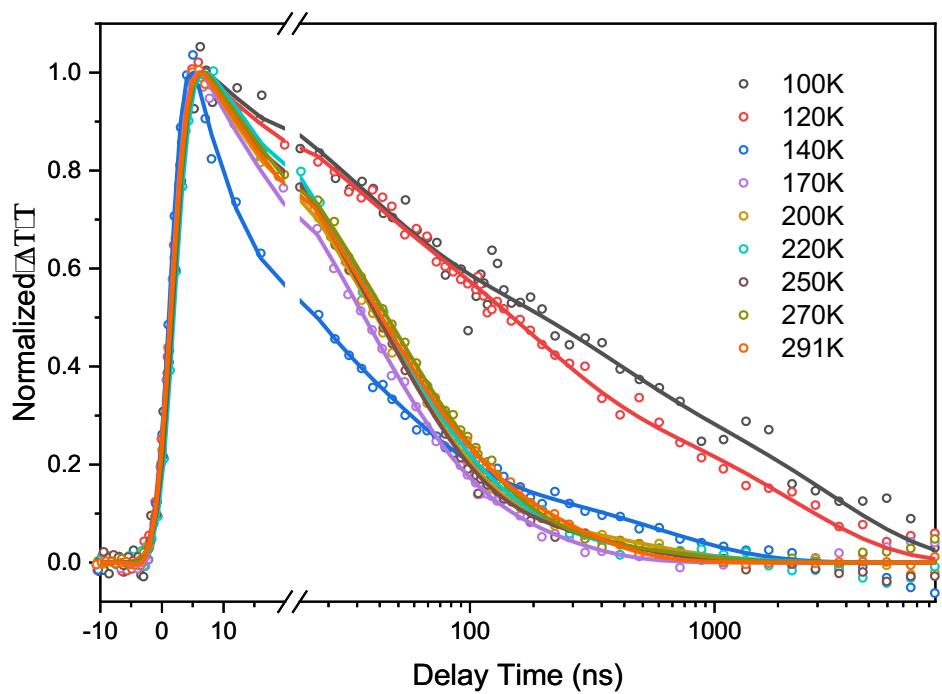

**Figure S9: Comparison of TA dynamics excited by nanosecond laser at different temperatures in the linear response range.** Dots are original data, and lines are fitting results.

| Temp. (K) | Res. (ps) | $A_1$ | $t_1$ (ps) | $A_2$ | $t_2$ (ps) | $A_3$ | $t_3$ (ns) |
|---|---|---|---|---|---|---|---|
| 100 | 0.28 ±0.00 | 0.690 ±0.014 | 0.796 ±0.034 | 0.143 ±0.001 | 13.5 ±1.8 | 0.330 ±0.005 | 17.1 ±3.3 |
| 120 | 0.27 ±0.01 | 0.609 ±0.017 | 0.718 ±0.035 | 0.144 ±0.001 | 28.3 ±4.2 | 0.416 ±0.001 | 16.4 ±3.1 |
| 140 | 0.30 ±0.01 | 0.830 ±0.044 | 0.411 ±0.034 | 0.265 ±0.012 | 13.2 ±1.2 | 0.270 ±0.008 | 5.2 ±0.8 |
| 170 | 0.36 ±0.01 | 0.949 ±0.053 | 0.242 ±0.013 | -0.039 ±0.005 | 95.8 ±34.2 | 0.715 ±0.006 | 38.2 ±6.9 |
| 200 | 0.36 ±0.01 | 0.752 ±0.033 | 0.263 ±0.011 | -0.033 ±0.003 | 84.0 ±23.7 | 0.763 ±0.004 | 62.3 ±11.5 |
| 220 | 0.36 ±0.00 | 0.699 ±0.029 | 0.265 ±0.010 | -0.043 ±0.003 | 110.2 ±23.2 | 0.804 ±0.004 | 42.4 ±4.9 |
| 250 | 0.36 ±0.00 | 0.646 ±0.029 | 0.278 ±0.011 | -0.046 ±0.004 | 171.4 ±35.9 | 0.823 ±0.005 | 37.1 ±4.2 |
| 270 | 0.37 ±0.00 | 0.625 ±0.026 | 0.284 ±0.011 | -0.041 ±0.004 | 144.7 ±31.9 | 0.824 ±0.004 | 44.1 ±5.4 |
| 291 | 0.36 ±0.00 | 0.448 ±0.021 | 0.334 ±0.015 | -0.039 ±0.010 | 418.1 ±168.0 | 0.867 ±0.011 | 47.3 ±11.8 |

**Table S2. Fitting results of TA curve in Figure S7 with triple exponential function.** $A_1$, $A_2$ and $A_3$ are the initial quantities of three decays, $t_1$, $t_2$ and $t_3$ are their lifetimes, and Res. is the resolution of TA system obtained from fitting. The data at 140 K are measured at the peak of 735 nm.

# Full OPTP dataset and analysis

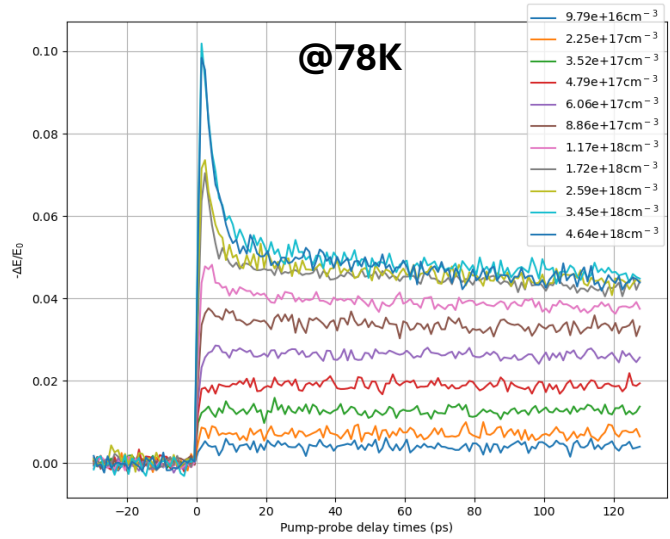
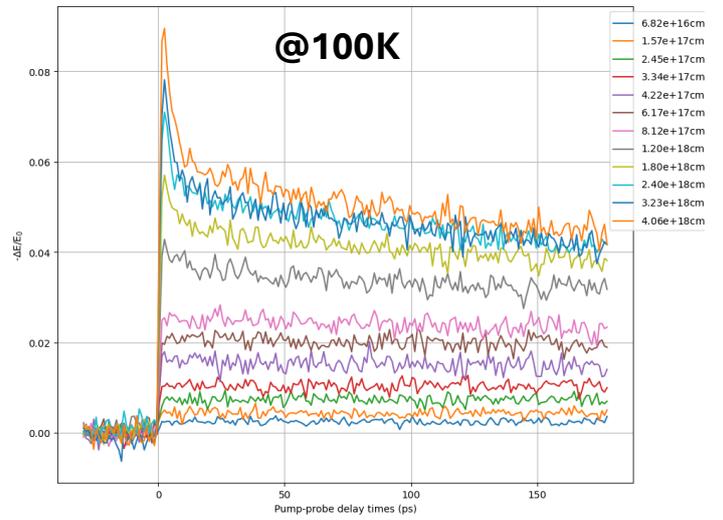
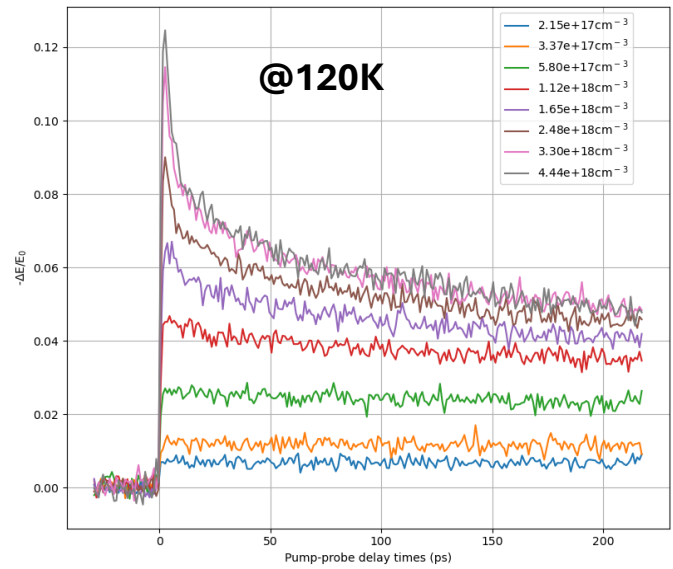
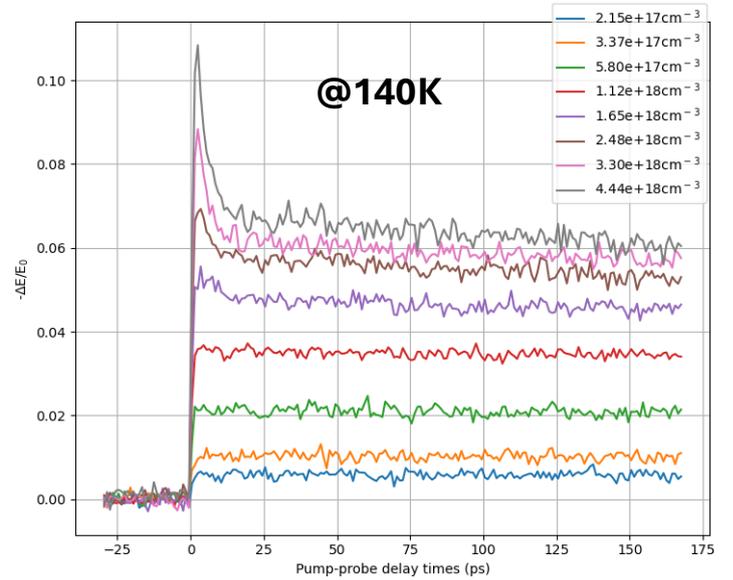
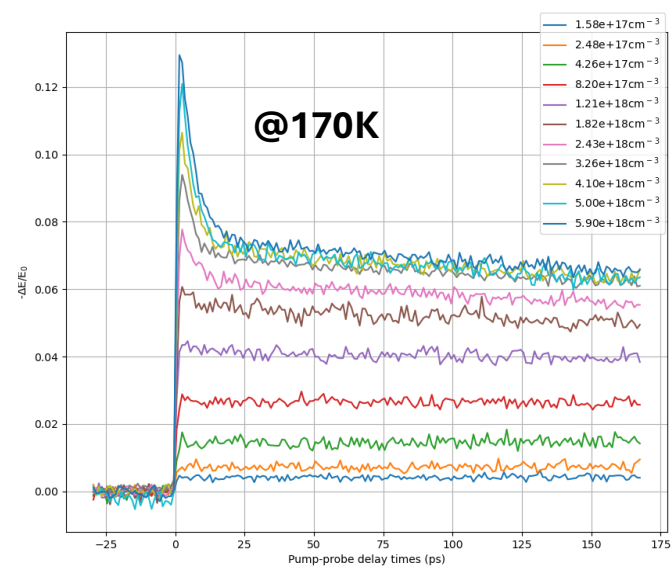
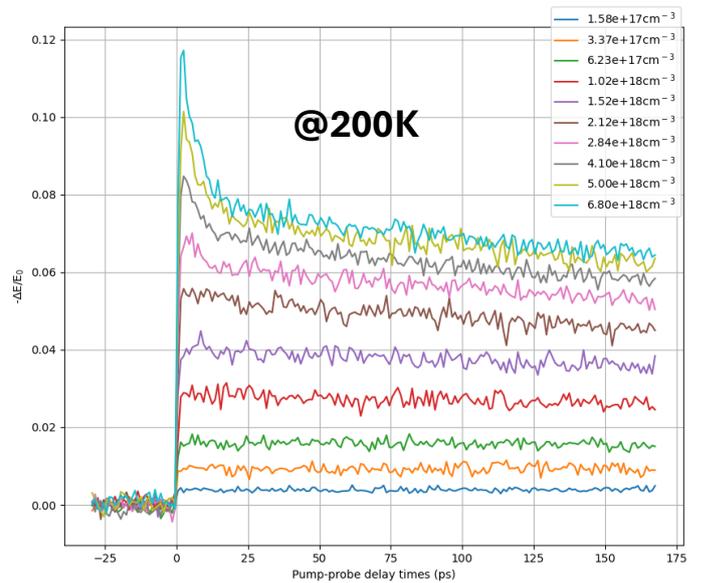

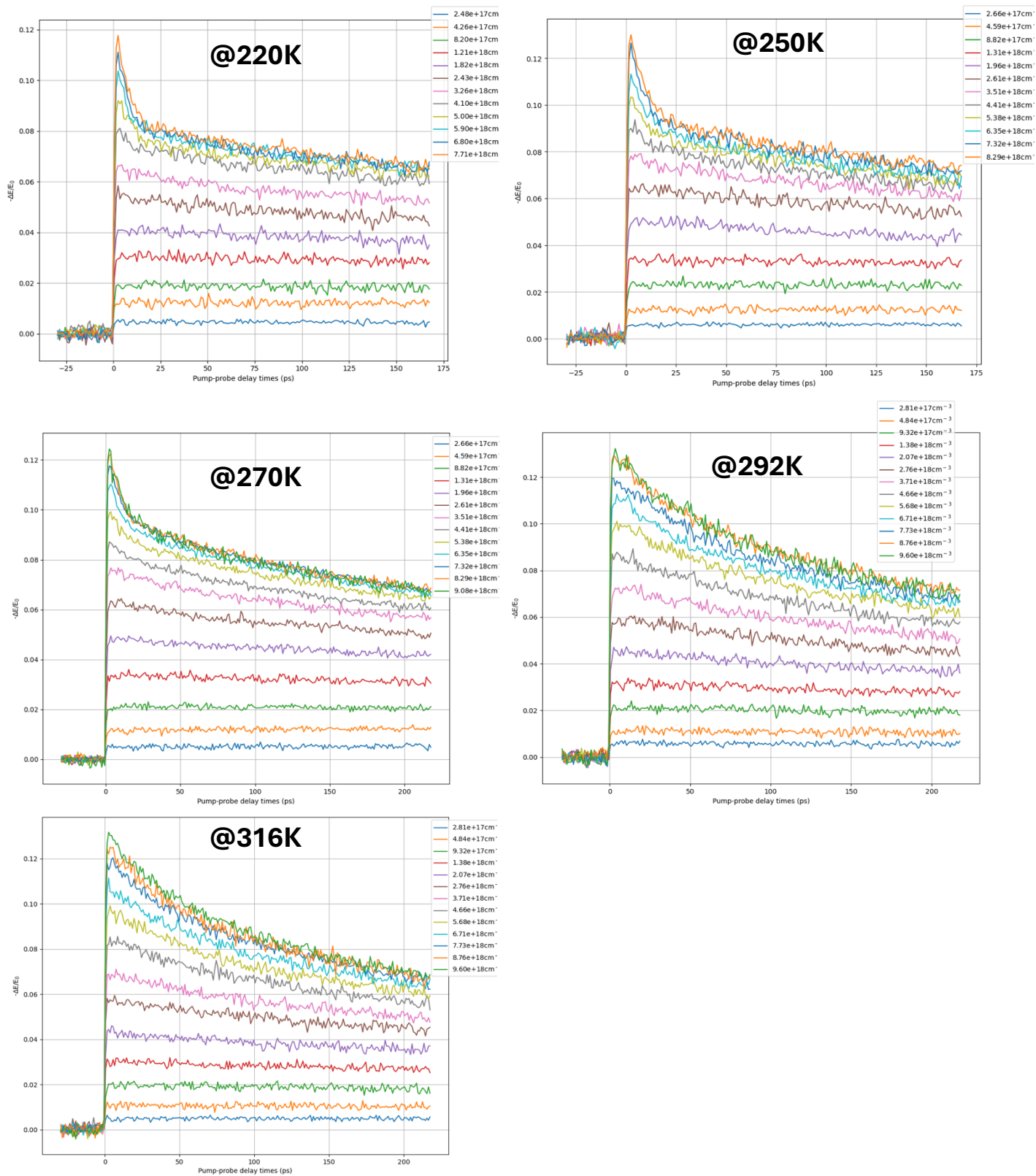

**Figure S10: Real part of the OPTP signal, by measuring the photoinduced change in the transmitted THz field -ΔE/E₀ (proportional to the photoconductivity) at the peak of the THz waveform.** Photoexcited carrier densities are indicated in the panel.

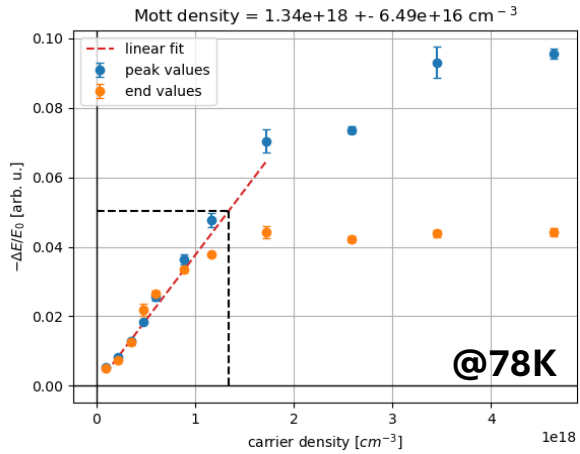
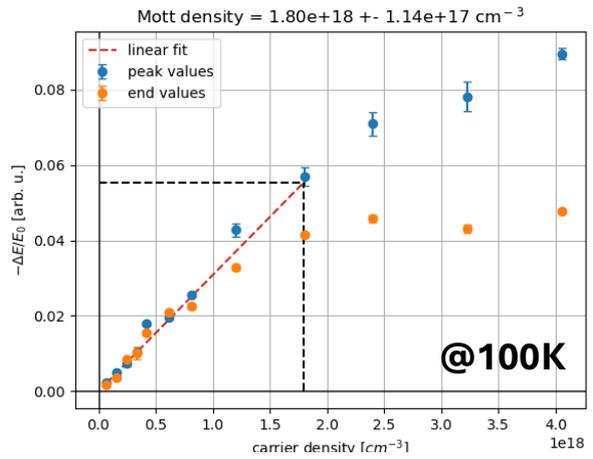
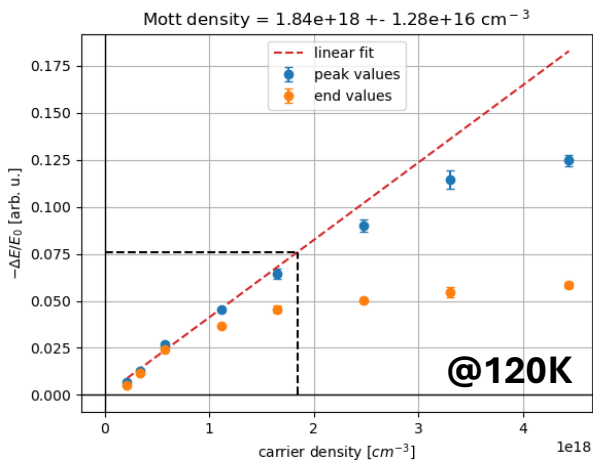
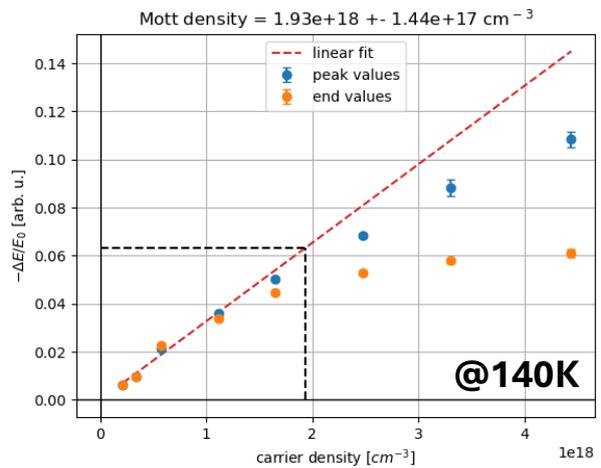
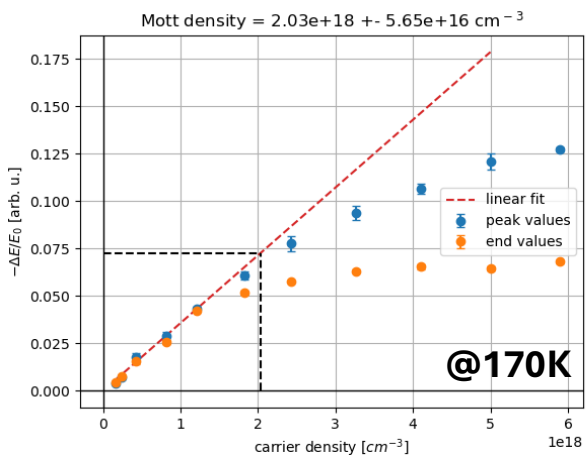
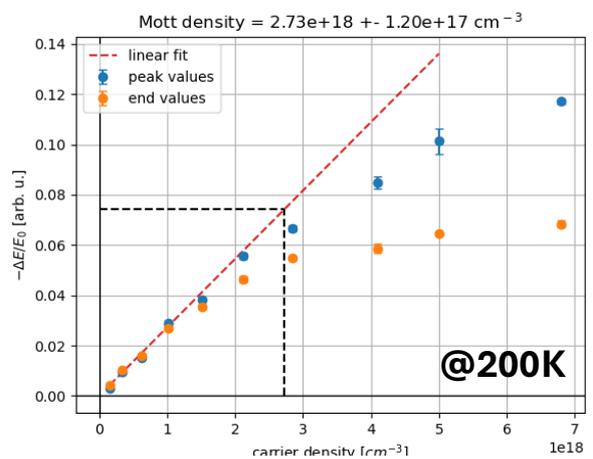
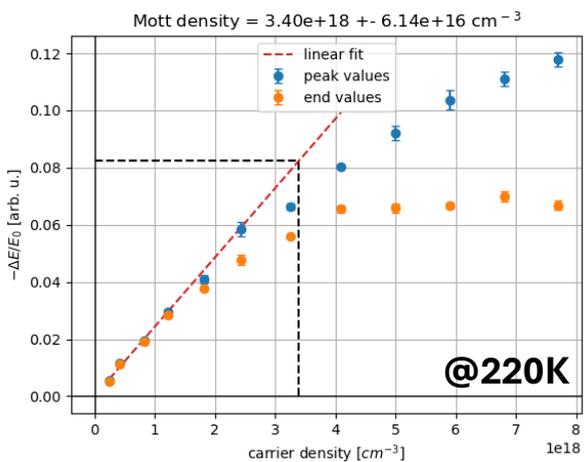
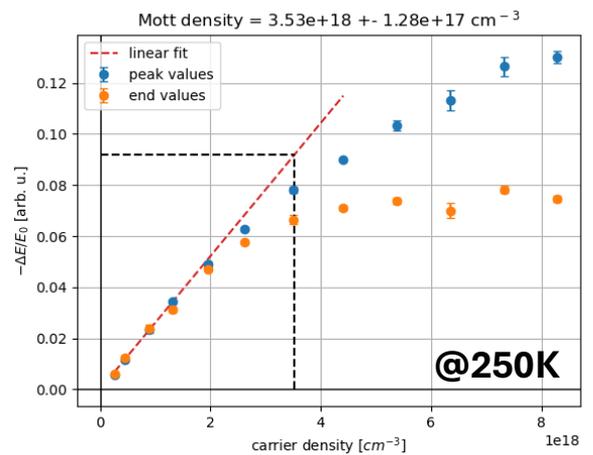

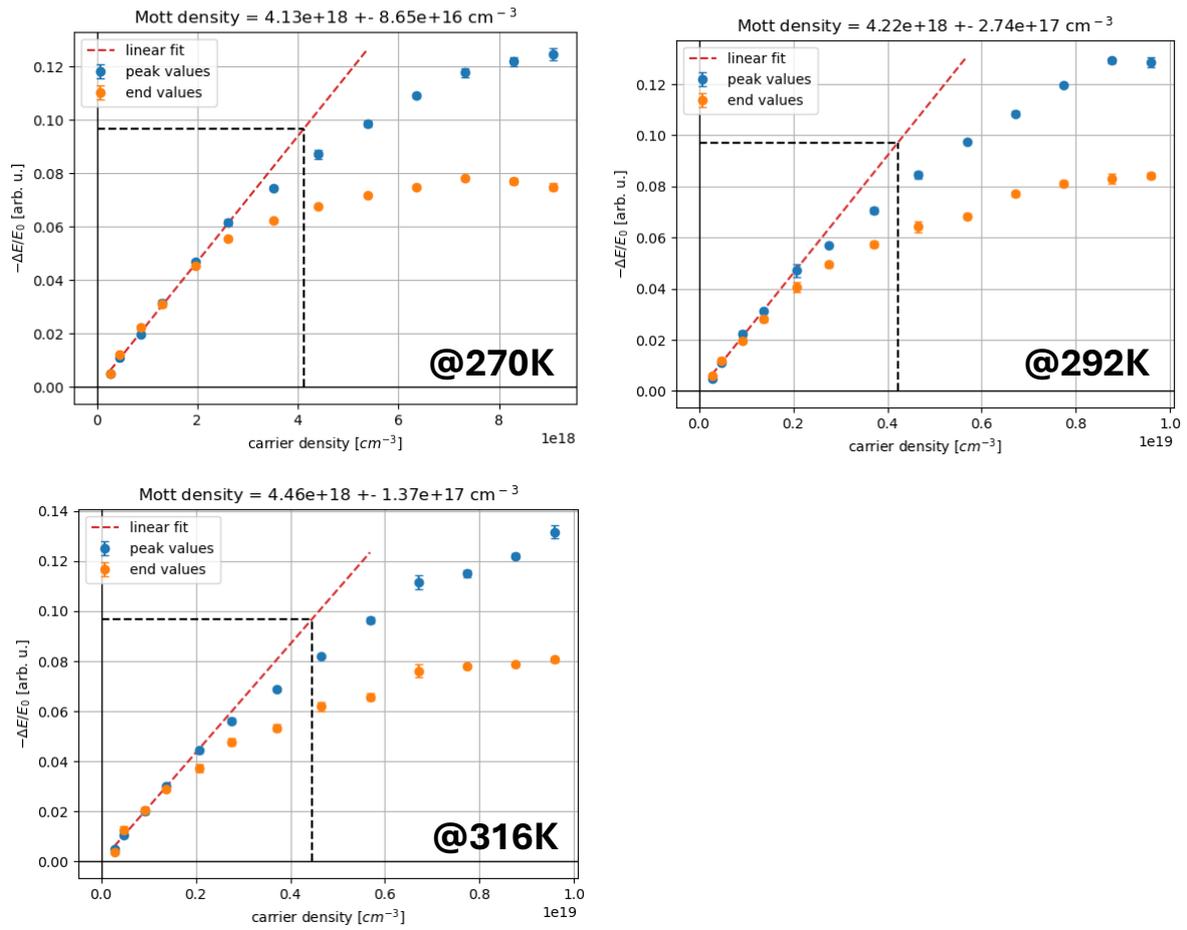

**Figure S11: Determination of the Mott density from the real part of the OPTP transients at various temperatures, acquired from the data shown in Figure SXX.** Blue datapoints are the $-\Delta E/E_0$ values at the peak of the OPTP signal (averaged from 0.5-1.5 ps pump-probe delay time). The yellow datapoints are the average $-\Delta E/E_0$ values over the last 10 ps of the OPTP transients. Error bars indicate standard deviations over the averaged time window. The red dashed line is a linear fit to the lowest three to five datapoints of the peak $-\Delta E/E_0$ value, where we have selected the range over which the increase of the signal was linear with increasing carrier density (i.e. pump fluence). We used the linear fit, as explained throughout the main text, to determine the Mott density at each temperature, which is indicated above each panel.

## Estimation of the temperature-dependent polaron size

At the Mott density, the polaron wavefunctions in MAPI start overlapping and annihilate rapidly. Assuming spherical wavefunctions, which occupy in a most-dense configuration where 74% of space is filled, we can state that the ratio of all volume occupied by polarons w.r.t. the total volume follows:

$$\frac{\#polarons \cdot \frac{4}{3}\pi r_{polaron}^3}{V} = 0.74$$

Where $r_{polaron}^3$ is the radius of one polaron cubed. We then use the definition of the Mott density $N_{mott}$ to calculate the radius of one polaron, which is converted into a diameter in Figure S5 below:

$$N_{mott} \cdot \frac{4}{3}\pi r_{polaron}^3 = 0.74$$

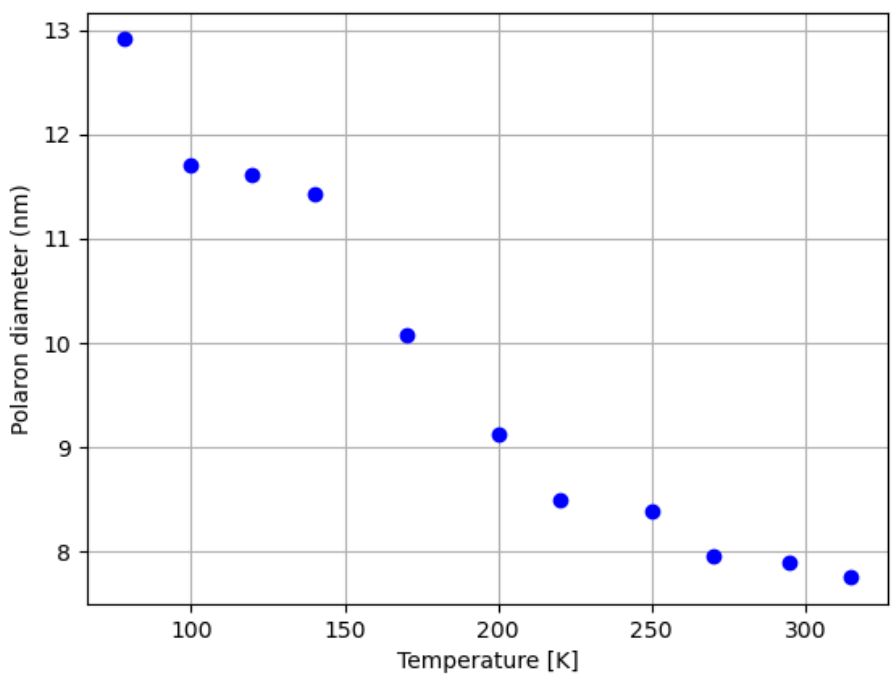

**Figure S12: Polaron diameter as a function of temperature, as obtained from the experimental OPTP data.**

As can be seen, the polaron diameter increases with decreasing temperature. Moreover, we confirm that the polaron wavefunctions span several unit cells (lattice parameters ~0.6 nm), showing that large polarons are formed in MAPI over the entire temperature range we probed experimentally.

## Modeling the threshold for optical gain/population inversion

We follow the same approach as Suárez et al.[3], which model at which density MAPI becomes transparent and shows optical gain, i.e. the density at which there is population inversion[4,5]. They follow the classic Bernard-Duraffourg conditions for lasing, namely that the splitting of quasi-Fermi levels (QFL) should be larger than or equal to the bandgap energy:

$$Ef_c - E_{f,v} \geq E_g$$

Or equivalently

$$Ef_c - E_{f,v} - E_g = 0$$

For this, we need to calculate how much a photoexcited electron and hole raise or lower the QFL towards the conduction band (for electrons) and valence band (for holes), respectively, for which we need to know the density of states. We use the density of a free electron gas in three dimensions in a parabolic band with effect mass *m\**:

$$N(E) = \int_{E_c}^{E} dE \; \frac{1}{2\pi^2} \left(\frac{2m_e^*}{\hbar^2}\right)^{3/2} \sqrt{E} \; \frac{1}{1 + e^{\frac{E_c - E_{f,c}}{k_B T}}}$$

With $m_e^*$ the effective mass of the first conduction band, $E_c$ the energy of the conduction band minimum, $E_{f,c}$ the QFL of the electrons and $N(E)$ the density of states per unit volume. The latter term in the integral is the Fermi-Dirac distribution for the electrons. A similar equation is set up for the holes. We calculate how much the addition of one electron (hole) raises (lowers) the QFL for the electrons.

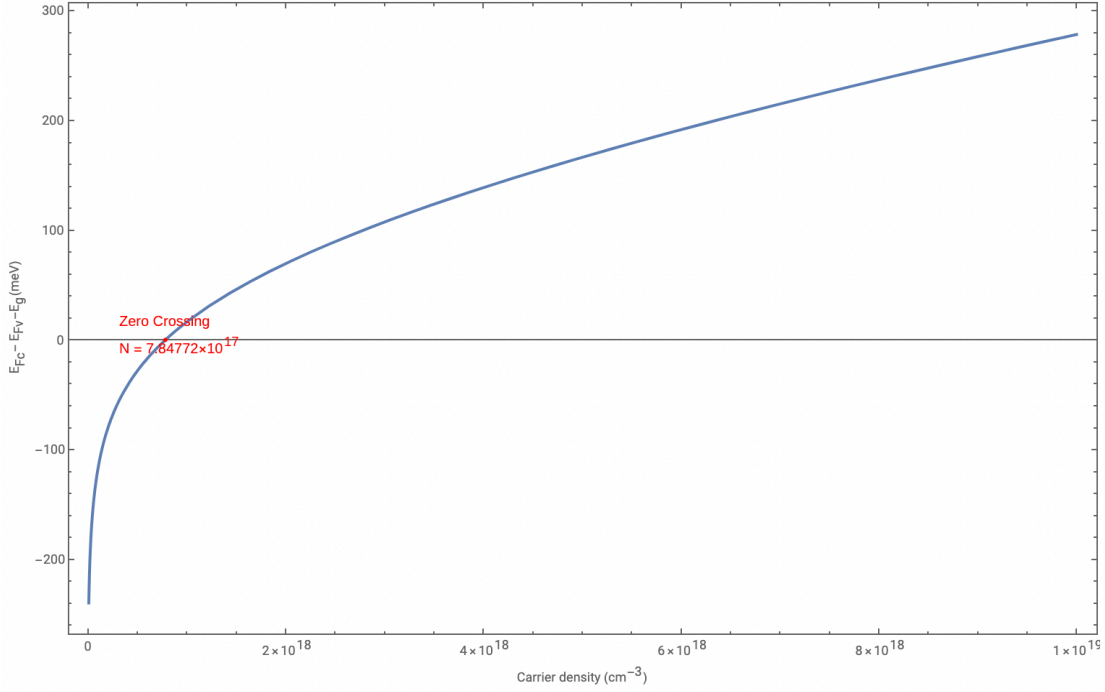

**Figure S13: example calculations of the threshold for optical gain in MAPI, with $m_e^*, m_h^* = 0.11, 0.13$.** We have changed both temperature and the electron and hole effective masses to obtain the temperature-and-density-dependent curves shown in Figure 3 of the main text.

We repeated the calculations to determine the density for population inversion for a range of temperatures, the results of which are shown in Figure 3 of the main text. We benchmarked the code and obtained the same results as Suárez et al.[3] for both MAPI and GaAs. Note that we do not consider excitonic effects, or fast-and-strong electron-phonon interactions, which would reduce the threshold for optical gain since the stimulated emission band gets redshifted out of the spectral range in which MAPI absorbs light.

# Supplementary references

(1) Leguy, A. M. A.; Hu, Y.; Campoy-Quiles, M.; Alonso, M. I.; Weber, O. J.; Azarhoosh, P.; van Schilfgaarde, M.; Weller, M. T.; Bein, T.; Nelson, J.; Docampo, P.; Barnes, P. R. F. Reversible Hydration of CH3NH3PbI3 in Films, Single Crystals, and Solar Cells. *Chem. Mater.* **2015**, *27* (9), 3397–3407. https://doi.org/10.1021/acs.chemmater.5b00660.

(2) Gao, L.; Zhang, H.; Zhang, Y.; Fu, S.; Geuchies, J. J.; Valli, D.; Saha, R. A.; Pradhan, B.; Roeffaers, M.; Debroye, E.; Hofkens, J.; Lu, J.; Ni, Z.; Wang, H. I.; Bonn, M. Tailoring Polaron Dimensions in Lead-Tin Hybrid Perovskites. *Advanced Materials n/a* (n/a), 2406109. https://doi.org/10.1002/adma.202406109.

(3) Suárez, I.; Juárez-Pérez, E. J.; Chirvony, V. S.; Mora-Seró, I.; Martínez-Pastor, J. P. Mechanisms of Spontaneous and Amplified Spontaneous Emission in ${\mathrm{CH}}_{3}{\mathrm{NH}}_{3}{\mathrm{Pb}}\mathrm{I}}_{3}$ Perovskite Thin Films Integrated in an Optical Waveguide. *Phys. Rev. Appl.* **2020**, *13* (6), 064071. https://doi.org/10.1103/PhysRevApplied.13.064071.

(4) Bernard, M. G. A.; Duraffourg, G. Laser Conditions in Semiconductors. *physica status solidi (b)* **1961**, *1* (7), 699–703. https://doi.org/10.1002/pssb.19610010703.

(5) Rosencher, E.; Vinter, B. *Optoelectronics*; Piva, P. G., Translator; Cambridge University Press: Cambridge, 2002. https://doi.org/10.1017/CBO9780511754647.